\documentclass[a4paper,fleqn,usenatbib]{mnras}
\usepackage[T1]{fontenc}
\usepackage{ae,aecompl}

\usepackage{graphicx}	% Including figure files
\usepackage{amsmath}	% Advanced maths commands
\usepackage{amssymb}	% Extra maths symbols
\usepackage{siunitx}

\usepackage[export]{adjustbox}

\usepackage{setspace}
\usepackage{etoolbox}
\makeatletter
\patchcmd\@combinedblfloats{\box\@outputbox}{\unvbox\@outputbox}{}{%
   \errmessage{\noexpand\@combinedblfloats could not be patched}%
}%
 \makeatother 
 
\newcommand{\be}{\begin{equation}}
\newcommand{\ee}{\end{equation}}
\newcommand{\bea}{\begin{eqnarray}}
\newcommand{\eea}{\end{eqnarray}}

\def\Sec#1{Section~\ref{sec:#1}}

\def\Fig#1{Fig.~\ref{fig:#1}}

\def\ifm#1{\relax\ifmmode#1\else$\mathsurround=0pt #1$\fi}
\def\kms{\ifmmode\,{\rm km}\,{\rm s}^{-1}\else km$\,$s$^{-1}$\fi}

\def\Msun{\,{\rm M_{\odot}}}
\def\Lsun{\,{\rm L_{\odot}}}

\def\Mstar{M_{\star}}

\def\ltsima{$\; \buildrel < \over \sim \;$}
\def\simlt{\lower.5ex\hbox{\ltsima}}
\def\gtsima{$\; \buildrel > \over \sim \;$}
\def\simgt{\lower.5ex\hbox{\gtsima}}

\newcommand{\lom}{LoM-50}
\newcommand{\him}{HiM-50}
\def\MLEsalp{\ifmmode\,{\rm MLE}_{\rm r, Salp}\else MLE$_{\rm r, Salp}$\fi}

\def\MLEkroupa{\ifmmode {\rm MLE}_{\rm r, Kroupa}\else MLE$_{\rm r, Kroupa}$\fi}

\def\r200{r_{200}}
\def\m200{m_{200}}
\def\V200{V_{200}}
\def\M200{M_{200}}
\def\R200{R_{200}}

\def\fdwarf{f_{\rm dwarf}}
\def\fdwarfSalp{f_{\rm dwarf, Salp}}

\newcommand{\ptm}{\vphantom}
\title[Variable IMFs with EAGLE -- III. Radial gradients]{Calibrated, cosmological hydrodynamical simulations with variable IMFs III: Spatially-resolved properties and evolution}
 
\author[C. Barber et al.]{
Christopher Barber,$^{1}$\thanks{Email: \href{mailto:cbar@strw.leidenuniv.nl}{cbar@strw.leidenuniv.nl}}
Joop Schaye$^{1}$,
and Robert A. Crain$^{2}$
\\
$^{1}$Leiden Observatory, Leiden University, PO Box 9513, NL-2300 RA Leiden, The Netherlands\\
$^{2}$Astrophysics Research Institute, Liverpool John Moores University, 146 Brownlow Hill, Liverpool L3 5RF, UK
}

\date{Accepted XXX. Received YYY; in original form ZZZ}

\pubyear{2018}

\graphicspath{{Figures_3/}}

\begin{document}
\label{firstpage}
\pagerange{\pageref{firstpage}--\pageref{lastpage}}
\maketitle

\begin{abstract}
%This is a simple template for authors to write new MNRAS papers.
%The abstract should briefly describe the aims, methods, and main results of the paper.
%It should be a single paragraph not more than 250 words (200 words for Letters).
%No references should appear in the abstract. 

Recent spatially-resolved observations of massive early-type galaxies (ETGs) have uncovered evidence for radial gradients of the stellar initial mass function (IMF), ranging from super-Salpeter IMFs in the centre to Milky Way-like beyond the half-light radius, $r_e$. We compare these findings with our new cosmological, hydrodynamical simulations based on the EAGLE model that self-consistently vary the IMF on a per-particle basis such that it becomes either bottom-heavy (LoM-50) or top-heavy (HiM-50) in high-pressure environments. These simulations were calibrated to reproduce inferred IMF variations such that the IMF becomes ``heavier'' due to either excess dwarf stars or stellar remnants, respectively, in galaxies with increasing stellar velocity dispersion. In agreement with observations, both simulations produce negative radial IMF gradients, transitioning from high to low excess mass-to-light ratio (MLE) at around $r_e$. We find negative metallicity radial gradients for both simulations, but positive and flat [Mg/Fe] gradients in LoM-50 and HiM-50, respectively. Measured in radial bins, the MLE increases strongly with local metallicity for both simulations, in agreement with observations. However, the local MLE increases and decreases with local [Mg/Fe] in LoM-50 and HiM-50, respectively. These qualitative differences can be used to break degeneracies in the manner with which the IMF varies in these high-mass ETGs. At $z=2$, we find that the MLE has a higher and lower normalization for bottom- and top-heavy IMF variations, respectively. We speculate that a hybrid of our LoM and HiM models may be able to reconcile observed IMF diagnostics in star-forming galaxies and ETGs.
\end{abstract}

\begin{keywords}
   methods: numerical -- stars: luminosity function, mass function  -- galaxies: structure -- galaxies: evolution -- galaxies: fundamental parameters -- galaxies: stellar content -- galaxies: elliptical and lenticular, cD.
\end{keywords}

\section{Introduction}

How early-type galaxies (ETGs) form and evolve over cosmic time is an area of active research, where one of the leading theories invokes an ``inside-out'' formation scenario that occurs in two distinct phases \citep{Bezanson2009, Oser2010, Barro2013, Clauwens2018}. First, the dense stellar core is formed at high redshift, in a rapid, high-pressure burst of star formation. In the second phase, stellar mass is accreted through minor and major mergers with other galaxies, adding material to the outer regions of ETGs. The different physical conditions under which the stars form in each of these phases gives rise to gradients in stellar properties as a function of radius \citep{Mehlert2003, Kuntschner2010, Greene2015}.

Equipped with the spatial and spectral resolving power of modern, panoramic integral field units, observational studies have recently inferred that the stellar initial mass function (IMF) varies radially in the centres of some local high-mass ETGs. In general, such studies conclude that the IMF is bottom-heavy in galaxy centres due to an excess of dwarf stars, transitioning to an IMF consistent with a Kroupa (i.e. Milky-Way-like) IMF from $\approx 0.1$ to 1 times the half-light radius, $r_e$ \citep{Boroson1991, Martin-Navarro2015b, LaBarbera2016, vanDokkum2017, Sarzi2018, Oldham2018a}. The presence of such gradients is still energetically debated, however, with some studies finding no gradients or arguing that it is currently too difficult to disentangle the IMF from radial abundance gradients \citep{McConnell2016, Zieleniewski2017, Davis2017, Alton2017, Alton2018, Vaughan2018a, Vaughan2018b}.

Such findings are important, since the IMF is usually assumed to be universal, both in the interpretation of observations and when making predictions with galaxy formation models. For example, radial IMF gradients imply that the stellar mass-to-light ratio in these galaxies varies as a function of radius, which could have a strong impact on dynamical mass models in which it is typically assumed that the $M/L$ ratios measured in the central regions of galaxies apply globally \citep[see][]{Bernardi2018b, Sonnenfeld2018, Oldham2018b}. Indeed, radial IMF variations may affect measurements of the IMF itself when inferring it within some fixed aperture.

Understanding how the IMF varies within a galaxy can give insights into the nature of the variation itself, both in terms of the physical origin of the variations, as well which part of the IMF is varying, and in which direction. For example, \citet{Martin-Navarro2015c} concluded that the spatially-resolved IMF correlates with local metallicity in 24 high-mass ETGs from the CALIFA survey, which could imply that metallicity may play an important role in shaping the IMF. This result is qualitatively consistent with the recent findings of \citet{vanDokkum2017} for 6 high-mass ETGs. Interestingly, these two studies assume very different parametrizations of IMF variations, steepening the low-mass and high-mass IMF slopes, respectively. Since most metals are produced by high-mass stars, the parametrization of the IMF variations is crucial to the predictions of galaxy formation models that attempt to invoke such IMF variations. For instance, if the high-mass slope correlates with metallicity, it is not clear if such correlations are due to a causation in either direction or a coincidence, since both quantities also scale strongly with radius.

That present-day high-mass ETGs formed the majority of their stars at high redshift leads to questions concerning how the IMF may have evolved over time. Indeed, observations of strongly star-forming galaxies at high redshift conclude that the high-mass slope of the IMF may need to be shallower to account for their H$\alpha$ equivalent widths \citep{Nanayakkara2017} or abundance ratios \citep{Zhang2018}. On the other hand, observations of present-day high-mass ETGs, which are the descendants of high-redshift starbursts, are typically found to have bottom-heavy stellar populations, with their stellar spectra indicating an excess of dwarf stars relative to a Milky Way-like IMF \citep[e.g. ][]{Conroy2012b, LaBarbera2013}. These apparently contradictory results could be evidence of a time dependence of IMF variations.  Indeed, different forms of IMF parametrization can lead to very different predictions of IMF-related observational diagnostics at the present day, even for a fixed mass-to-light ratio \citep[][hereafter Paper I]{Barber2018a}. These differences may be even stronger at high redshift when the stars are actually forming.

Such issues can be addressed with galaxy formation models that explicitly include IMF variations. The recent cosmological, hydrodynamical simulations of Paper I are currently the most well-suited to answer these questions. The IMF variations in these simulations were assumed to depend on the local pressure of star-forming gas and were calibrated to reproduce the correlation between galaxy-wide mass-to-light excess and central stellar velocity dispersion, $\sigma_e$, inferred by \citet{Cappellari2013b}. This match was achieved by increasing the contribution of either low-mass dwarf stars or stellar remnants (black holes, neutron stars, and white dwarfs) to the stellar $M/L$ by varying the low-mass or high-mass IMF slope, respectively, to become bottom- or top-heavy in high-pressure environments on a per-particle basis. The IMF variations in these simulations are fully self-consistent in terms of the local star formation law, stellar energetic feedback, and nucleosynthetic yields.

We showed in Paper I that these simulations, which employ the EAGLE model for galaxy formation \citep{Schaye2015}, agree with the observables used to calibrate this model, namely the present day galaxy luminosity function, half-light radii, and supermassive black hole masses. In \citet[][hereafter Paper II]{Barber2018b} we investigated correlations between the ``galaxy-wide'' IMF and galaxy properties at $z=0.1$, including metallicity, [Mg/Fe], age, stellar mass, luminosity, size, and black hole mass. In this paper, the third in this series, we use our variable IMF simulations (summarized in \Sec{simulations}) to investigate if radial IMF gradients, which were not considered in the initial IMF calibrations, are predicted by our models (\Sec{IMF_vs_r}), and see how plausible IMF variations may affect radial gradients in stellar population properties such as the $M/L$ ratio, metallicities, and $\alpha$-enhancement (\Sec{starprops_gradients}). We also investigate how the local IMF varies with local properties, and directly compare with recent observations (\Sec{IMF_vs_local_starprops}). In \Sec{redshift} we investigate the evolution of the IMF and its effect on the evolution of galaxies in the simulations. We discuss our models and possible extensions in \Sec{Discussion}. Finally, we present our conclusions in \Sec{conclusions}. The simulation data is publicly available at \url{http://icc.dur.ac.uk/Eagle/database.php} \citep{McAlpine2016}.

\section{Simulations}
\label{sec:simulations}

%fig 1 - IMFs
\begin{figure*}
\includegraphics[width=0.45\textwidth]{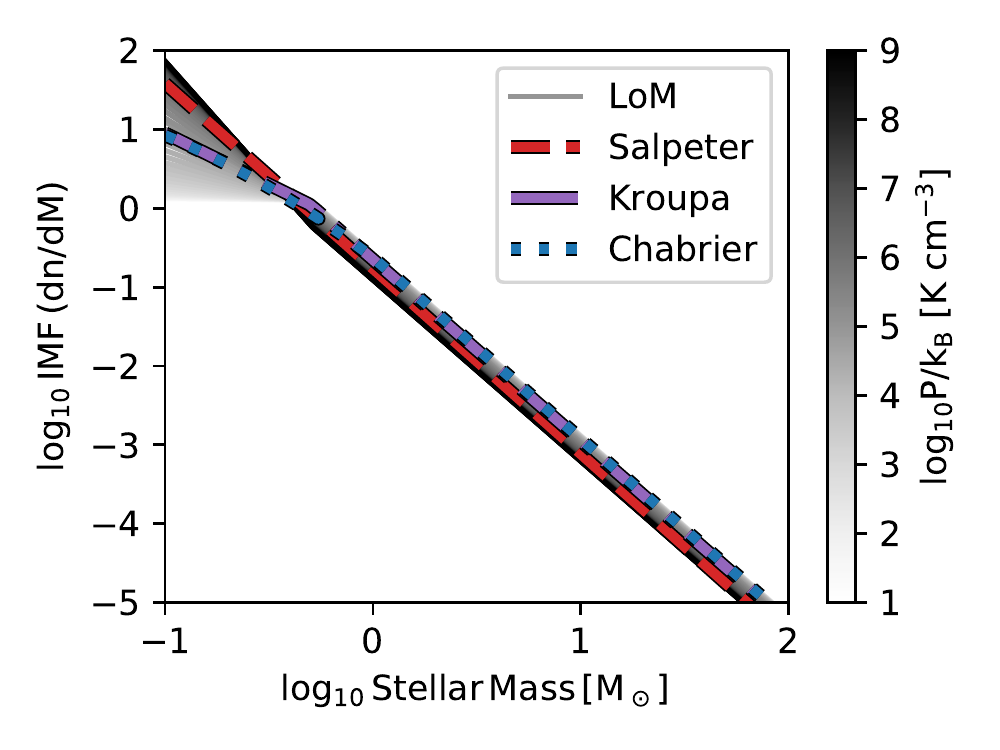}
\includegraphics[width=0.45\textwidth]{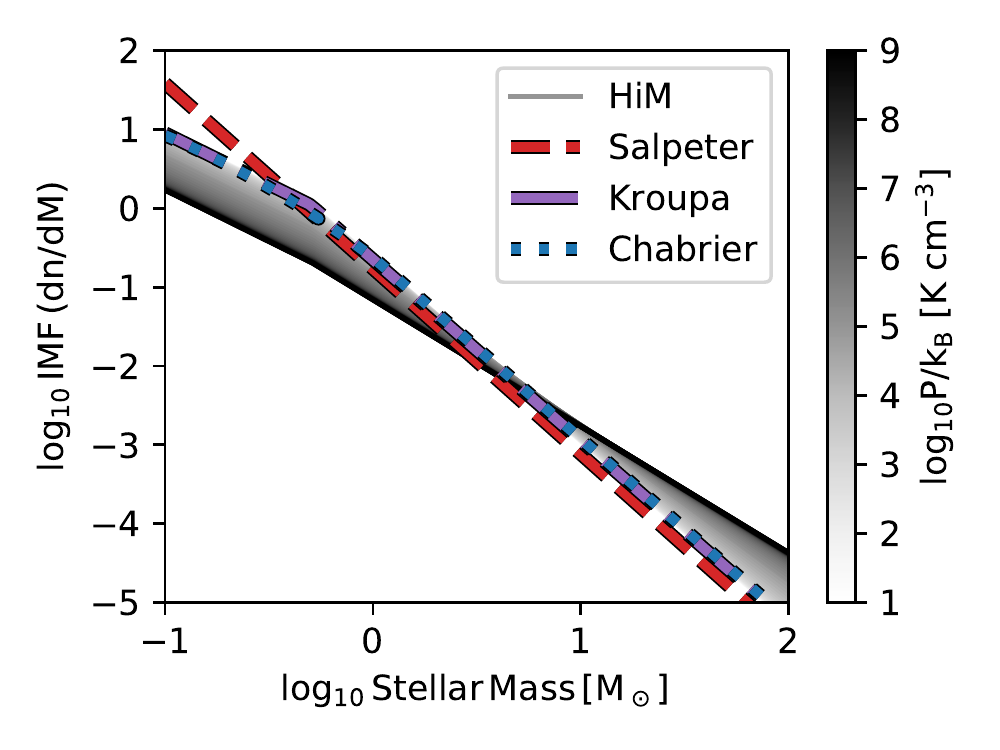}
\caption{IMF variation prescriptions employed by \lom{} (left panel) and \him{} (right panel). Grey lines show the IMF assigned to stellar populations for a range in birth ISM pressures (see greyscale bar). For all IMFs the integrated mass is normalized to $1 \Msun$. In \lom{}, the low-mass ($m < 0.5 \Msun$) IMF slope is varied such that the IMF transitions from bottom-light to bottom-heavy from low- to high-pressure environments. In \him{}, the high-mass ($m>0.5\Msun$) IMF slope is instead varied to become top-heavy in high-pressure environments. Figure reproduced from Fig. 2 of Paper I.}
\label{fig:IMF}
\end{figure*}

% fig 2 - radial variations, maps
 \begin{figure*}
   \centering
 \includegraphics[width=\textwidth]{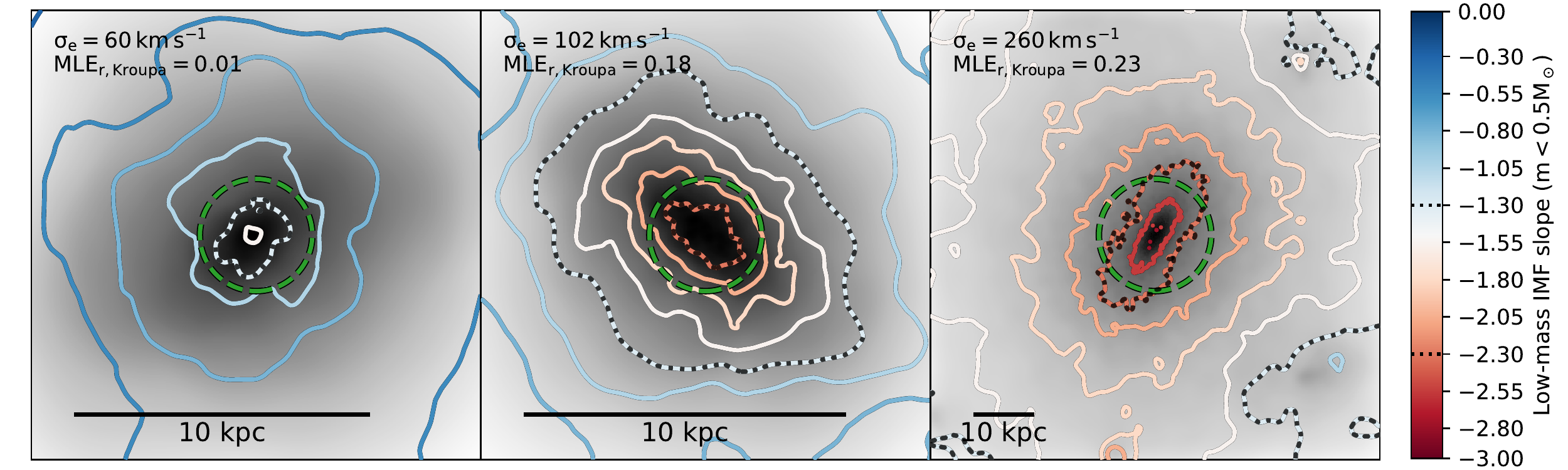}
  \includegraphics[width=\textwidth]{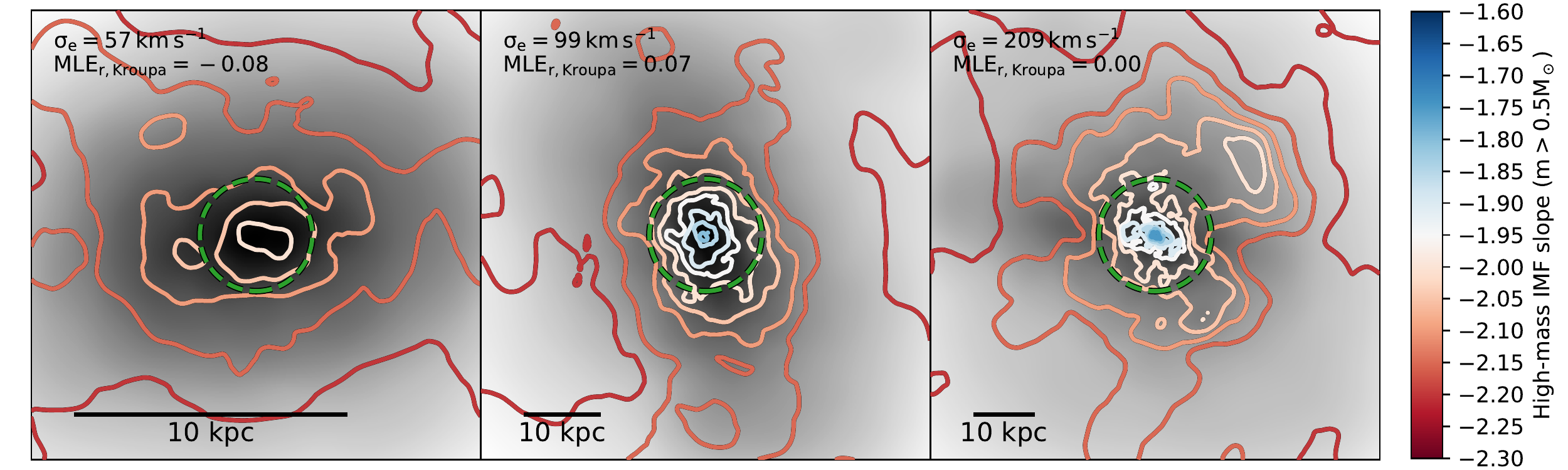}
 \caption{ IMF maps for three example galaxies of different masses from each of \lom{} (top row) and \him{} (bottom row) at $z=0.1$. Greyscale maps show logarithmic projected stellar mass surface density. The extent of each image is 8 times the 2D projected SDSS $r$-band half-light radii, $r_e$, which is marked by a dashed green circle in each panel. Physical proper kpc are indicated with scale bars.  Coloured contours (corresponding to ticks in the colour bars) denote mass-weighted IMF slope for $m < 0.5 \Msun$ (top row) and $m > 0.5 \Msun$ (bottom row). Bluer and redder colours correspond to shallower and steeper IMF slopes, respectively. In the upper row we highlight the low-mass IMF slope contours corresponding to Kroupa and Salpeter IMFs as dotted white and red lines, respectively, where applicable. The stellar velocity dispersion, $\sigma_e$, and excess mass-to-light ratio relative to a Kroupa IMF (\MLEkroupa{}, Equation \ref{eqn:MLEkroupa}), both $r$-band weighted and measured within $r_e$, are indicated in the upper left corner of each panel. Note that the upper and lower rows do not correspond to the same galaxies. \lom{} (\him{}) galaxies with higher $\sigma_e$ tend to have more bottom-heavy (top-heavy) IMFs in their central regions, with steep gradients of increasing heaviness toward their centres. Lower-mass galaxies have Chabrier-like IMFs and weaker IMF gradients.}
  \label{fig:alpha_maps}
 \end{figure*}

The simulations used in this work are a variation on the EAGLE project, a suite of cosmological, hydrodynamical simulations of galaxy formation and evolution \citep{Schaye2015, Crain2015, McAlpine2016}. They were run using the Tree-Particle-mesh smooth particle hydrodynamics code P-Gadget-3 \citep[last described by][]{Springel2005} in a periodic, comoving volume of (50 Mpc)$^{3}$ from $z=127$ to $z=0$. They have the same resolution as the fiducial EAGLE model, with particle mass of $m_{\rm g} = 1.8 \times 10^6 \Msun$ and $m_{\rm DM} = 9.7 \times 10^6 \Msun$ for gas and dark matter, respectively. The gravitational softening length was set to 2.66 co-moving kpc for $z > 2.8$ and held fixed at 0.70 proper kpc thereafter. A Lambda cold dark matter cosmogony is assumed, with cosmological parameters chosen for consistency with Planck 2013 \citep[$\Omega_{\rm b} = 0.04825$, $\Omega_{\rm m} = 0.307$, $\Omega_{\Lambda}=0.693$, $h = 0.6777$;][]{Planck2014}.

In the reference EAGLE model, physical processes that occur below the resolution limit of the simulation are modelled via  ``sub-grid'' prescriptions. Gas particles cool radiatively according to cooling rates determined for the 11 individually-tracked elements that are most important for cooling \citep{Wiersma2009a}, and are photoheated by an evolving, spatially uniform \citet{Haardt2001} UV/X-ray background and the cosmic microwave background. When gas particles become sufficiently dense and cool \citep{Schaye2004}, they become eligible to stochastically transform into star particles. These star particles evolve and lose mass via supernovae (SNe) type Ia and II, and winds from AGB stars and massive stars \citep{Wiersma2009b} according to an IMF (which in the reference EAGLE model is assumed to universally follow a \citealt{Chabrier2003a} form) and metallicity-dependent lifetimes of \citet{Portinari1998}. Thermal stellar feedback is implemented stochastically \citep{DallaVecchia2012} and was calibrated to match the $z\approx0$ galaxy stellar mass function (GSMF) and galaxy sizes. Supermassive black holes are seeded in high-mass haloes and can grow via BH-BH mergers or gas accretion \citep{Springel2005a, Booth2009, Rosas-Guevara2015}, the latter leading to thermal AGN feedback that quenches star formation in high-mass galaxies. We refer the reader to \citet{Schaye2015} for a more detailed description of the EAGLE model and to \citet{Crain2015} for details of its calibration.

The models employed in our two variable IMF simulations (first presented in Paper I) differ from the reference EAGLE model in that, rather than assuming a fixed Chabrier IMF for all stellar populations, they self-consistently vary the IMF for individual star particles as a function of the pressure of the interstellar medium (ISM) in which each star particle forms\footnote{We note also that on kpc scales the pressure correlates well with the star formation rate surface density through the Kennicutt-Schmidt law \citep[e.g.][]{Schaye2008}, so our IMF prescription can also be interpreted as implementing a dependence of the IMF on other variables such as the radiation or cosmic ray density.}.  The IMF is defined over the range $0.1 - 100 \Msun$ with a mass-dependent slope $x(m)$ as ${\rm d}n/{\rm d}m \propto m^{x(m)}$. The two IMF variation prescriptions  studied are shown in \Fig{IMF}. The left panel shows the model termed ``LoM'', where the IMF transitions from bottom-light to bottom-heavy from low to high pressures by varying the low-mass ($m < 0.5 \Msun$) slope of the IMF while keeping the high-mass slope fixed at the Kroupa value of $x=-2.3$. For the second prescription, called ``HiM'' (right panel), we instead vary the high-mass slope ($m > 0.5 \Msun$) to transition from Kroupa-like to top-heavy from low to high pressures while keeping the low-mass slope fixed at the Kroupa value of $x=-1.3$. Due to the finite resolution of EAGLE, this pressure corresponds to the ISM pressure averaged on scales of $\approx$ 1 kpc. Note that for a self-gravitating disc, variations with pressure are equivalent to variations with star formation rate (SFR) surface density \citep{Schaye2008}. Crucially, both IMF variation prescriptions were calibrated to broadly reproduce the correlation between the excess mass-to-light ratio relative to that expected given a fixed IMF (MLE), and central stellar velocity dispersion, $\sigma_e$, inferred for high-mass ETGs by \citet{Cappellari2013b}. Stellar feedback, the star formation law, and metal yields were all modified to self-consistently account for the local IMF variations. We refer the reader to Paper I for further details of these IMF prescriptions and their calibration. 

Since the simulations do not explicitly model the emission of optical light, we compute photometric luminosities for all stellar particles (in post-processing) using the flexible stellar population synthesis (FSPS) software package \citep{Conroy2009, Conroy2010}, using the Basel spectral library \citep{Lejeune1997, Lejeune1998, Westera2002} with Padova isochrones \citep{Marigo2007, Marigo2008}, where the stellar population's age, metallicity, initial stellar mass, and IMF are all taken into account. For consistency we also recompute stellar masses using FSPS, and note that for the highest-mass ($\Mstar>10^{11}\Msun$) galaxies in \him{} this leads to larger stellar masses of up to 0.2 dex due to differences in how stellar remnants are computed in FSPS and the \citet{Wiersma2009b} model built into EAGLE. We assume that stellar populations with ${\rm age} < 10\,$Myr have zero luminosity since such populations are expected to be obscured by their birth clouds \citep{Charlot2000}, but otherwise account for no further effects of dust. 

Galaxies are identified in the simulations using a friends-of-friends halo finder combined with the SUBFIND algorithm which identifies self-bound structures \citep{Springel2001,Dolag2009}. We ensure that all galaxies studied in this work are well-sampled, with each containing at least 500 bound stellar particles. Unless otherwise stated, we compute global galaxy properties such as the stellar velocity dispersion, $\sigma_e$, as a Sloan Digital Sky Survey (SDSS) $r$-band light-weighted average over all bound star particles within the 2D projected $r$-band half-light radius, $r_e$, of each galaxy, where the projection is along the ``random'' $z$-axis of the simulation. 

In the upper row of \Fig{alpha_maps} we show random projections of three example galaxies taken from \lom{} at $z=0.1$ with $\sigma_e \approx 60$, 100, and $>150\kms$. The greyscale background shows projected stellar mass density maps, while coloured contours show maps of the mass-weighted low-mass ($m<0.5\Msun$) IMF slope. For the high-$\sigma_e$ galaxies, we see strong radial gradients of the IMF slope, starting as Chabrier-like in the outskirts and becoming more bottom-heavy towards the centre. Some lower-$\sigma_e$ galaxies also show IMF gradients, but these are much weaker than those with higher $\sigma_e$. Analogously, in the lower row of \Fig{alpha_maps} we show the same but for the \him{} simulation (with a different set of galaxies). Here we also see strong radial gradients in the highest-$\sigma_e$ galaxies, but now the IMF becomes more top-heavy toward the centre. 

\begin{figure}
   \centering
\includegraphics[width=0.45\textwidth]{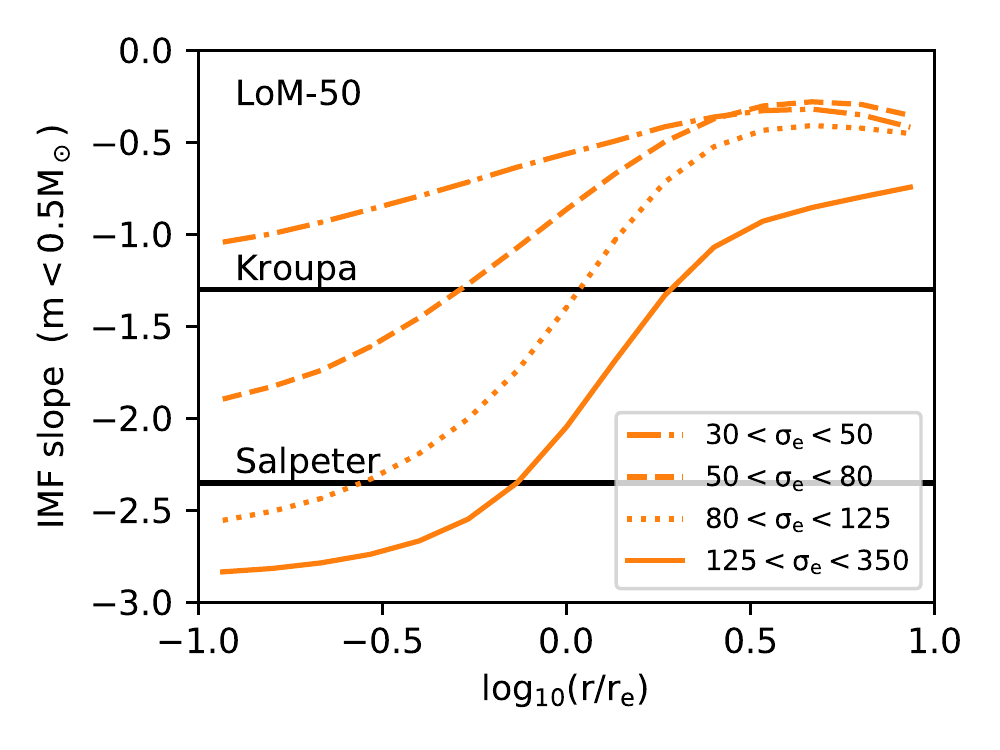}
\includegraphics[width=0.45\textwidth]{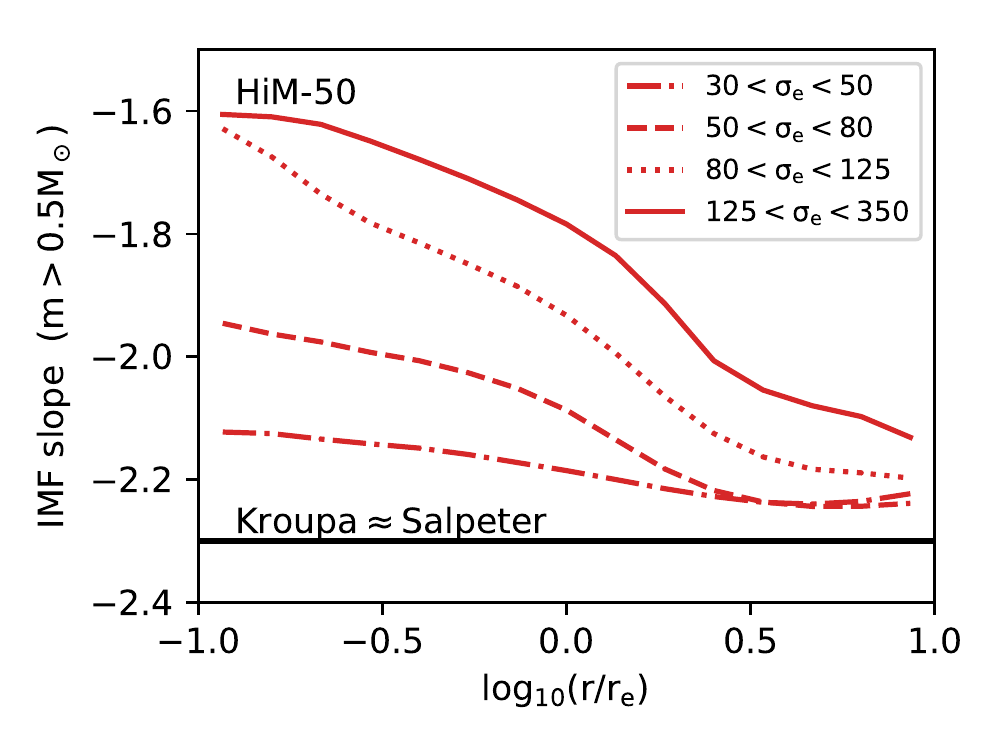}

 \caption{IMF slope as a function of 2d projected galactocentric radius at $z=0.1$, normalized to the half-light radius of each galaxy. Coloured lines show median values for all stars belonging to subhaloes with stellar velocity dispersion, $\sigma_e$, in bins defined by the legend (in units of $\kms$). The upper panel shows the low-mass ($m<0.5\Msun$) slope for \lom{} galaxies while the lower panel shows the high-mass ($m>0.5\Msun$) slope for \him{} galaxies. IMF slopes for Salpeter and Kroupa IMFs are indicated with horizontal solid lines. In both simulations IMF gradients are stronger in higher-$\sigma_e$ galaxies.}
  \label{fig:IMF_vs_r_sigmabins}
 \end{figure}

We show these trends statistically in \Fig{IMF_vs_r_sigmabins}, where in the upper and lower panels we plot respectively for \lom{} and \him{} the radial profiles of the median IMF slope over the region of the IMF that is varied for star particles belonging to subhaloes in bins of $\sigma_e$. We see here that while high-$\sigma_e$ galaxies ($ \sigma_e > 125 \kms$) have strong radial IMF gradients, low-$\sigma_e$ galaxies ($30 < \sigma_e/\kms < 50$) have relatively flat profiles, remaining Kroupa-like at all radii. In the next section we explore these radial trends in high-$\sigma_e$ galaxies in detail and compare with observations.

%fig 3: sigma dependence of radial IMF slope profiles

\section{IMF trends within galaxies }
\label{sec:IMF_internal}

In this section we present the spatially-resolved properties of high-$\sigma_e$ galaxies in our variable IMF simulations. \Sec{sample} describes the sample of high-mass ETGs selected for comparison with observations, for which we investigate radial gradients in the IMF and some of its observational diagnostics in \Sec{IMF_vs_r}. \Sec{starprops_gradients} shows the effect of IMF variations on gradients in stellar properties, and in \Sec{IMF_vs_local_starprops} we present the resulting correlations between the local IMF and stellar properties within individual galaxies. 

\subsection{Sample selection}
\label{sec:sample}

 %fig 4 - veldisp selection
 \begin{figure}
   \centering
 \includegraphics[width=0.5\textwidth]{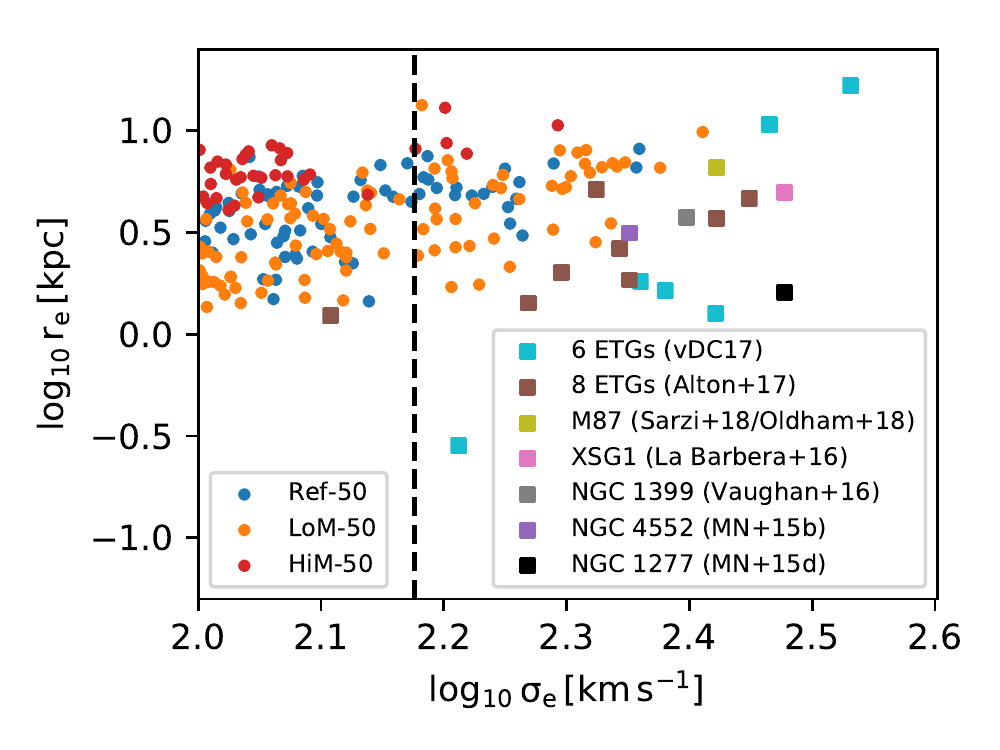}
 \caption{ Projected proper $r$-band half-light radius, $r_e$, as a function of projected $r$-band light-weighted stellar velocity dispersion measured within $r_e$, $\sigma_e$, for early-type galaxies (ETGs; defined as those with intrinsic $u^*-r^*>2$) in Ref-50 (dark blue dots), \lom{} (orange dots) and \him{} (red dots) at $z=0.1$. We compare with observed ETGs with recently measured spatially-resolved IMFs shown as squares: the 6 high-$\sigma_e$ ETGs studied by \citet{vanDokkum2017} in cyan, eight ETGs studied by \citet{Alton2017} in brown, M87 in yellow (studied separately by \citealt{Sarzi2018} and \citealt{Oldham2018a}), XSG1 in pink \citep{LaBarbera2016}, NGC 1399 in grey \citep{Vaughan2018b}, NGC 4552 in purple \citep{Martin-Navarro2015b}, and NGC 1277 in black (\citealt{Martin-Navarro2015d}, values from \citealt{VandenBosch2012}). Where possible, distances were taken from \citet{Cappellari2011} and $\sigma_e$ and $r_e$ from the above-mentioned references or \citet{Cappellari2013a}. For each simulation we select ETGs with $\sigma_e > 150\kms$ (hereafter the ``Sigma150'' sample) for comparison with observations, indicated by the vertical dashed line. Our Sigma150 galaxies have on average lower $\sigma_e$ than the observed samples, which should be kept in mind when comparing our results with observations.}
  \label{fig:sample}
 \end{figure}

In order to compare the internal properties of galaxies in our variable IMF simulations with observations, we select ETGs with $\sigma_e > 150\kms$, hereafter referred to as the ``Sigma150'' sample. As in Papers I and II, we define ETGs as those with intrinsic (dust-free) $u^*-r^*>2$. The $\sigma_e$ limit of $150\kms$ was chosen as a compromise between having a statistically significant number of galaxies in our sample and fairness of comparison with observational samples of high-$\sigma$ galaxies. This selection leaves us with 40 and 5 Sigma150 galaxies in \lom{} and \him{}, respectively. The difference in sample sizes is due to a combination of lower typical $\sigma_e$ values and lower passive fractions in \him{} galaxies (see Paper I). Note that selecting all galaxies with $\sigma_e>150\kms$ rather than only ETGs would increase our sample sizes for \lom{} and \him{} to 67 and 12 respectively, but would make no qualitative difference to any of the results presented in this paper.

\Fig{sample} shows $r_e$ as a function of $\sigma_e$ for our simulated ETGs, and compares with various observed high-$\sigma$ ETGs that have spatially-resolved IMF measurements. These observed samples will be described in \Sec{IMF_vs_r}. Our Sigma150 galaxies are slightly larger and have lower $\sigma_e$ on average compared to the observational sample. However, note that for three of the observed galaxies from \citealt{vanDokkum2017}, including the smallest one, $r_e$ is measured along the projected semi-minor axis, which may be smaller than our (circularly-averaged) $r_e$ values. These differences should be kept in mind when comparing IMF diagnostics with observed galaxies in future sections.

\subsection{Radial IMF gradients}
\label{sec:IMF_vs_r}

To measure the radial dependence of the IMF within our galaxies, it is important to account for the non-circularity of the galaxy surface brightness profiles (see \Fig{alpha_maps}), as is also done in observational studies \citep[e.g.][]{Parikh2018}. To do so, we fit an ellipse, with semi-major axis\footnote{This choice is motivated by some observational IMF studies that measure $r_e$ along the semi-major axis of the galaxy \citep[e.g.][]{Martin-Navarro2015d, vanDokkum2017}. For ETGs, the difference between the circularly-averaged $r_e$ and the $r_e$ measured along the semi-major axis is usually small, around 0.1 dex (see Table 1 of \citealt{Cappellari2013a}), so this choice has little impact on our results.} equal to the circularly-averaged $r_e$, to the 2D projected surface brightness profile of each galaxy. We then measure the $r$-band light-weighted IMF slope for each galaxy within concentric elliptical shells by scaling this ellipse such that its semi-major axis is logarithmically-spaced in bins of width 0.1 dex, ranging from $\log_{10} r/r_e = -1$ to $1$. We caution that, given the gravitational softening length of 0.7 kpc and that $r_e$ for most of our galaxies lies in the range 2-10 kpc, any radial gradients at $r \lesssim 0.1-0.3\,r_e$ are likely affected by numerical resolution or the equation of state imposed on star-forming gas in EAGLE \citep{Benitez-Llambay2018}. Therefore we show radial profiles only for $r > 0.1\,r_e$. Additionally, radii larger than 10$\,r_e$ would be well beyond what can be measured observationally. Indeed, due to the fact that spectroscopic IMF studies require a signal-to-noise ratio of at least 100 to detect variations in the dwarf-to-giant ratio, most such studies measure the IMF within, at most, a couple of $r_e$ \citep[e.g.][]{Vaughan2018a}.

In this section we explore radial gradients for various IMF-dependent diagnostics, including the IMF slope (\Sec{IMFslope_grads}), mass-to-light excess (\Sec{MLE_grads}), and the dwarf-to-giant ratio (\Sec{fdwarf_grads}).

% fig 5 - radial variations of IMF
\begin{figure}
   \centering
 \includegraphics[width=0.5\textwidth]{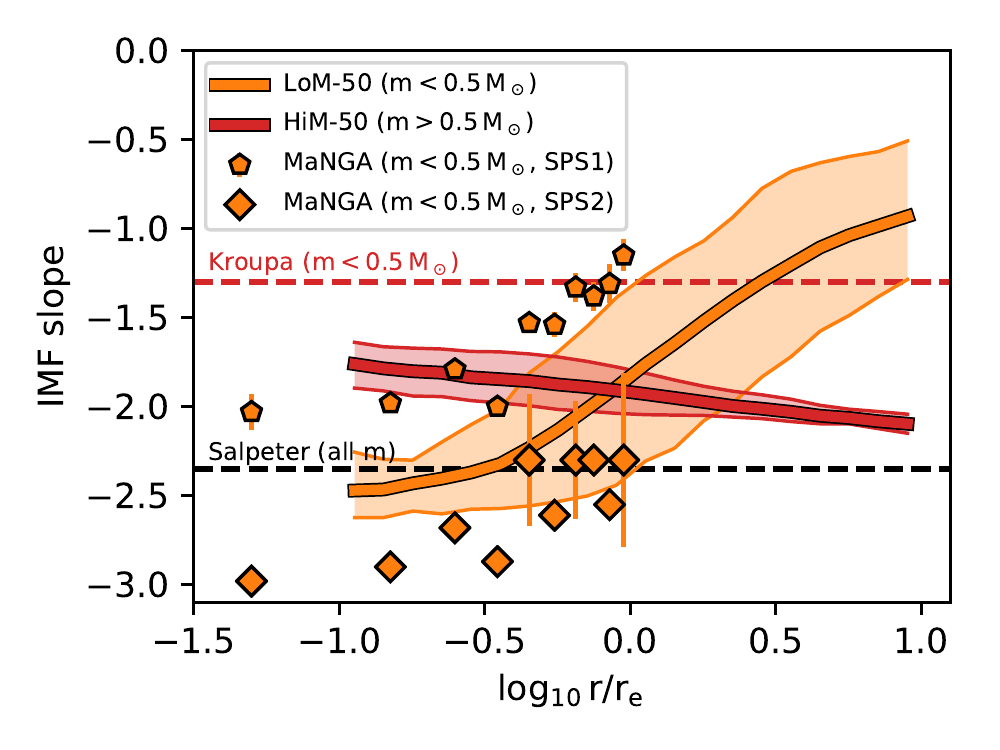}
 \caption{Radial variations in the $r$-band light-weighted IMF slope for ETGs with $\sigma_e>150\kms$ in the \lom{} (orange) and \him{} (red) simulations at $z=0.1$. We show only the slope for the mass range of the IMF that is varying: $m<0.5\Msun$ and $m>0.5\Msun$ for \lom{} and \him{}, respectively. Solid thick lines show mean values averaged over all galaxies within logarithmically-binned 2D concentric elliptical shells with semi-major axis $r/r_e$, while solid filled regions show 10-90$^{\rm th}$ percentiles. The dashed-red line marks the low-mass slope for a Kroupa IMF (which is also the low-mass slope for all galaxies in \him{}), while the black-dashed line marks the Salpeter slope at all masses. We compare \lom{} with low-mass slope variations for 122 ETGs in the mass range $\log_{10}\Mstar/\Msun \in [10.5,10.8]$ from SDSS-MaNGA as orange points where pentagons and diamonds correspond to differences in stellar population synthesis (SPS) modelling \citep{Parikh2018}. In both simulations we find strong radial IMF gradients, with the IMF becoming more bottom- and top-heavy toward the centres of galaxies for \lom{} and \him{}, respectively. Our low-mass slope variations in \lom{} agree well with observations within the errors associated with SPS modelling.}
  \label{fig:IMFslope_vs_r}
 \end{figure}
 
\subsubsection{IMF slope radial gradients} 
\label{sec:IMFslope_grads}

\Fig{IMFslope_vs_r} shows the IMF slope as a function of $r/r_e$ for our Sigma150 samples in \lom{} and \him{} at $z=0.1$, where we show the (varying) low-mass and high-mass slopes for the respective simulations. For both simulations we see strong radial gradients, with the IMF becoming bottom- and top-heavy toward the centre for \lom{} and \him{}, respectively.

We compare our \lom{} simulation with observed gradients in the low-mass IMF slope ($m<0.5\Msun$) inferred by \citet{Parikh2018} from NaI absorption features in the radially binned stacked spectra of 122 morphologically-selected ETGs from the SDSSIV Mapping Nearby Galaxies at APO (MaNGA) survey with $\log_{10}\Mstar/\Msun \in [10.5,10.8]$. To demonstrate the systematic uncertainty in the determination of the IMF slope via spectroscopic modelling, we show their results using two different stellar population synthesis (SPS) models: SPS1 uses the stellar population models of optical Lick absorption indices of \citet{Thomas2011a} in combination with \citet{Maraston2011} models based on the theoretical MARCS \citep{Gustafsson2008} library, while SPS2 uses the \citet{Villaume2017a} extended NASA Infrared Telescope Facility (IRTF) stellar library. While both models yield significant radial gradients in the IMF slope, they are significantly offset from one another. Encouragingly, they seem to straddle our \lom{} results, making the agreement between our simulations and these observations very good within the systematic uncertainties of the SPS modelling. 

Note that \him{}, which by construction has a low-mass slope fixed at the Kroupa value (dashed red line in \Fig{IMFslope_vs_r}), is not consistent with these observations. Indeed, observational SPS studies that parametrize IMF variations by varying the high-mass IMF slope typically find that it becomes steeper toward the centre \citep[e.g.][]{Martin-Navarro2015b, LaBarbera2016}, which is also in conflict with the \him{} simulation. However, such studies are only sensitive to the low-mass end of the IMF since only the long-lived stars (with $m\lesssim 1\Msun$) remain present in the old ETGs that are the focus of these IMF studies. Studies that are sensitive to the high-mass end of the IMF \citep[e.g.][]{Gunawardhana2011} and which inspired our HiM IMF variation prescription, have not yet explored radial gradients of the high-mass slope within galaxies. It will thus be an important test of the \him{} model to establish whether observations that are sensitive to the high-mass end of the IMF find its slope becomes shallower toward the centres of high-mass galaxies. 

% fig 6 - age and pressure vs r
\begin{figure}
\centering
\includegraphics[width=0.5\textwidth]{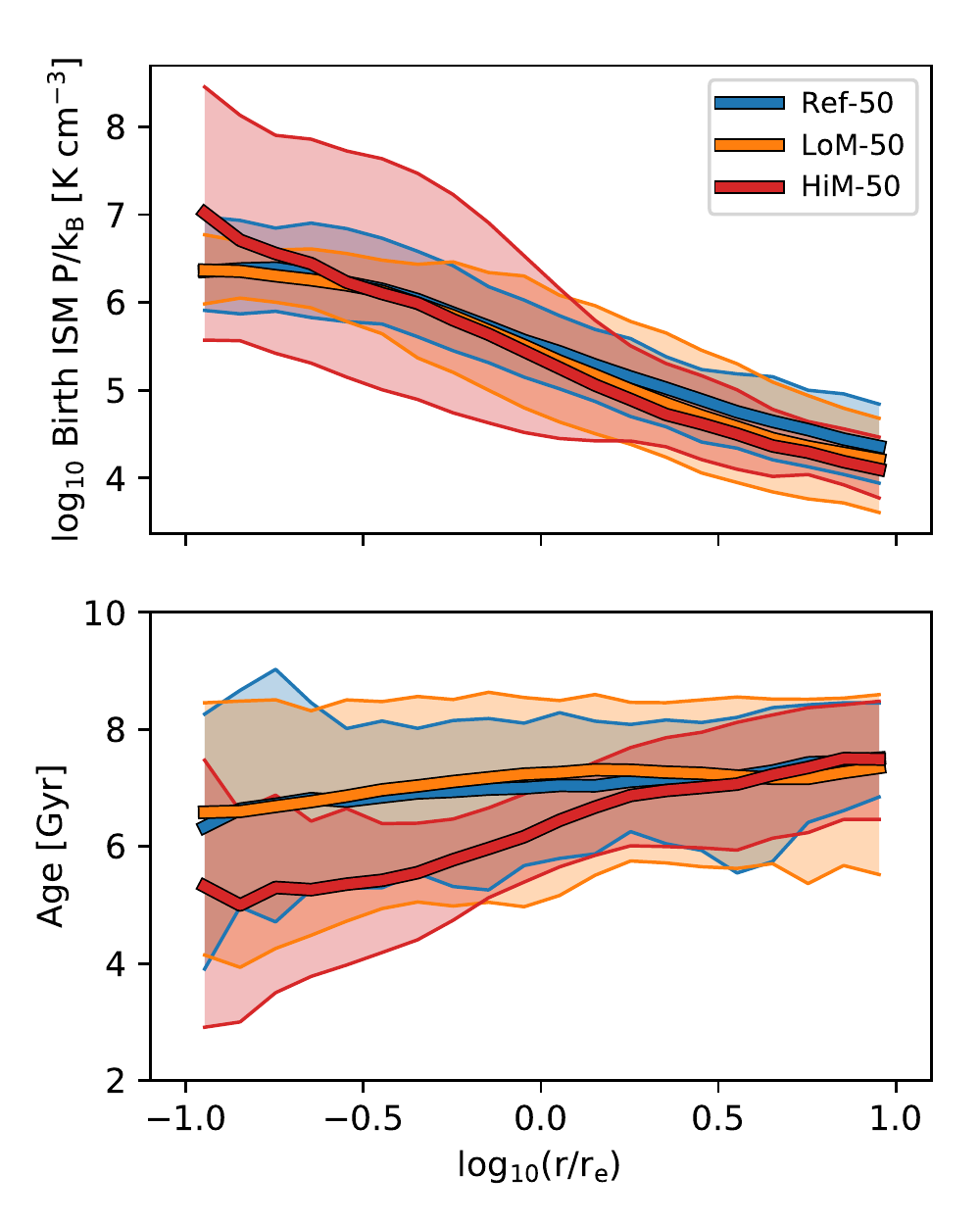}
 \caption{Radial variations of spatially-resolved properties of ETGs with $\sigma_e > 150\kms$ in Ref-50 (blue), \lom{} (orange), and \him{} (red) at $z=0.1$.  We show $r$-band light-weighted birth ISM pressure and stellar age in the upper and lower panels, respectively, within logarithmically-binned 2D projected concentric elliptical shells with semi-major axis $r/r_e$ (note that we use the locations of the stars at $z=0.1$, not at their formation time). Thick lines show values averaged over all galaxies, while filled regions show 10-90$^{\rm th}$ percentiles. The average birth pressure profiles are not very different between the three simulations, except with more scatter in \him{} than in \lom{} or Ref-50. The age profiles for Ref-50 and \lom{} are similar, but stars in \him{} galaxies are younger by $\approx 1-2$ Gyr for $r<r_e$}
  \label{fig:age_pressure_vs_r}
 \end{figure}

 % fig 7 - radial variations of MLE
 \begin{figure}
   \centering
 \includegraphics[width=0.5\textwidth]{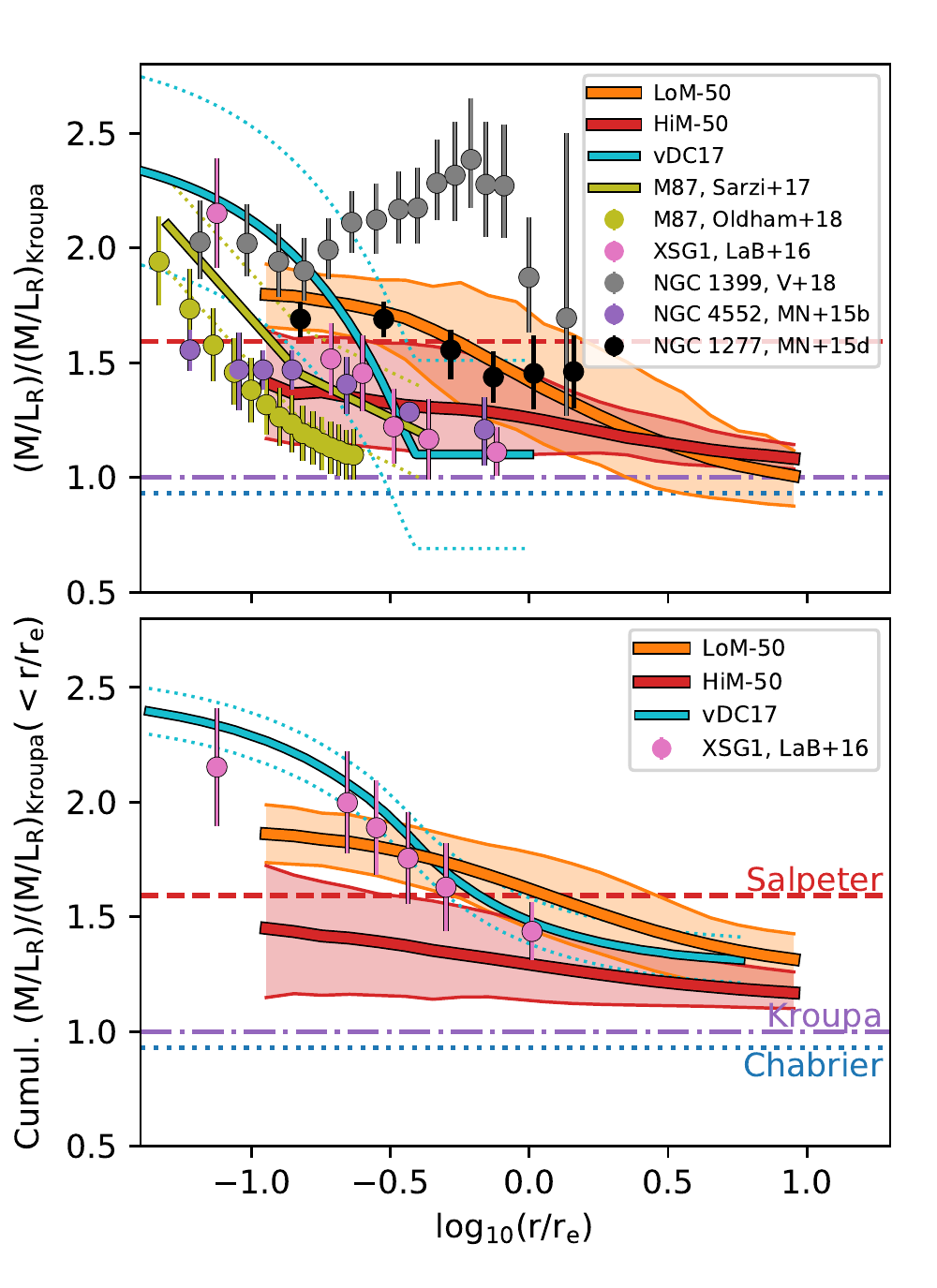}
 \caption{Radial variations in the $r$-band mass-to-light excess relative to a Kroupa IMF, \MLEkroupa{}, within ETGs with $\sigma_e > 150\kms$ in the \lom{} (orange) and \him{} (red) simulations at $z=0.1$. Radii are normalized to their $r$-band half-light radii, $r_e$. Solid thick lines show mean values averaged over all galaxies in each $r/r_e$ bin, while solid filled regions show 10-90$^{\rm th}$ percentiles. In the upper panel we show \MLEkroupa{} measured within logarithmically-binned 2D projected concentric elliptical shells with semi-major axis $r/r_e$, while the lower panel shows mean luminosity-weighted \MLEkroupa{} within circular 2D projected apertures of radius $r/r_e$. Radial MLE gradients for various observed ETGs are over-plotted (see legend and text). Horizontal red-dashed, purple-dot-dashed, and blue-dotted lines mark the expected values for universal Salpeter, Kroupa, and Chabrier IMFs, respectively. For both \lom{} and \him{}, the IMF is heavy in the centres and transitions to Kroupa-like at a few times $r_e$, qualitatively consistent with observations (although note that \him{} agrees with the spectroscopic observational studies for the wrong reasons; see \Fig{fdwarf_vs_r}).}
  \label{fig:MLE_vs_r}
 \end{figure}

  % fig 8 - radial variations of fdwarf and F05,1
 \begin{figure}
   \centering
 \includegraphics[width=0.48\textwidth]{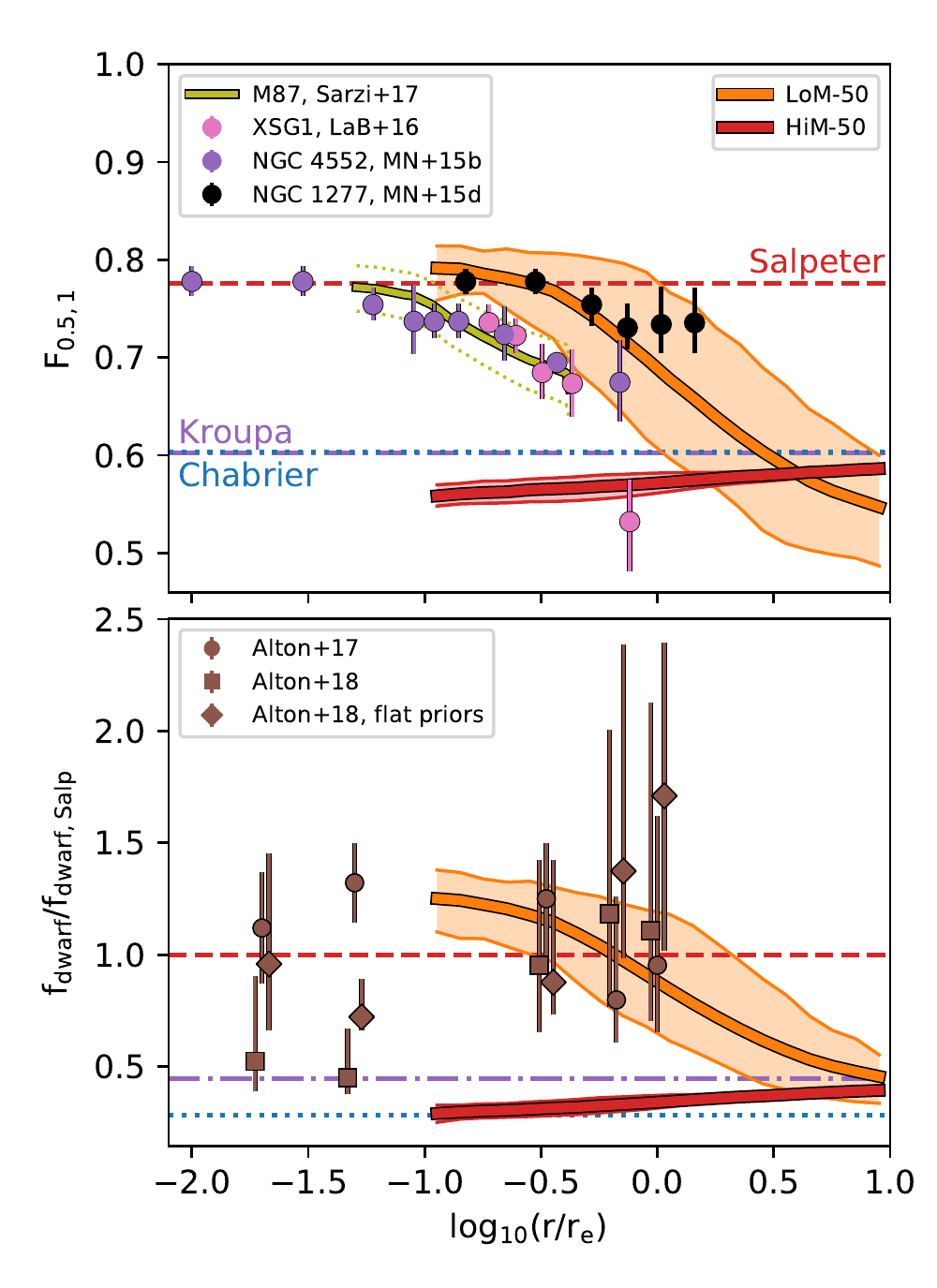}
 \caption{As \Fig{MLE_vs_r} but now showing dwarf-to-giant ratio radial profiles. In the upper panel we define this ratio as the mass fraction of stars with $m<0.5\Msun$ relative to those with $m<1\Msun$ in the IMF, $F_{0.5,1}$. In the lower panel, this ratio is defined as the fraction of the total $r$-band luminosity contributed by low-mass ($m<0.5\Msun$) stars at $z=0.1$, $f_{\rm dwarf}$. Due to the dependence of this quantity on the age of the stellar population, in each bin we divide by the value computed for a fixed Salpeter IMF, given the same ages and metallicities of the stars. Both quantities are measured within logarithmically-binned 2D concentric elliptical shells of semi-major axis $r/r_e$. In the upper panel we compare with $F_{0.5,1}$ profiles found for M87 (yellow line), XSG1 (pink points), NGC 4552 (purple points), and NGC 1277 (black points). In the lower panel we show the average $f_{\rm dwarf}$ for stacked spectra of 8 ETGs from \citet{Alton2018} as brown squares with error bars. Brown diamonds show their results assuming a flat prior in $\fdwarf$, while brown circles show their results for the same galaxies using older SPS models and fewer absorption features \citep{Alton2017}. Horizontal red-dashed, purple-dot-dashed, and blue-dotted lines mark the expected values for universal Salpeter, Kroupa, and Chabrier IMFs. While \lom{} galaxies match the observed dwarf-to-giant ratio radial gradients, the near-constant Kroupa/Chabrier values for \him{} galaxies are in tension with the observations.}
  \label{fig:fdwarf_vs_r}
 \end{figure}

These strong radial IMF gradients are a consequence of similar gradients in birth ISM pressure, which we show explicitly in the upper panel of \Fig{age_pressure_vs_r} for our Sigma150 galaxies\footnote{Note that the profiles presented in \Fig{age_pressure_vs_r} use the locations of the stars at $z=0.1$, rather than at their formation, and thus may be subject to radial migration \citep[e.g. Fig. 9 of][]{Furlong2017}. However, since radial migration affects all other $z=0.1$ radial profiles presented in this work in the same way, showing these pressure and age profiles at $z=0.1$ allows us to interpret our IMF trends.}. For all of our simulations, including Ref-50, birth ISM pressure increases by over 2 orders of magnitude from the outskirts to the centre. Interestingly, \him{} exhibits much more scatter in the central $r < 0.3\,r_e$ than in either the Ref-50 or \lom{} simulations, which translates into the greater scatter in the IMF slope in \him{} than in \lom{} (relative to the range over which it is varied in each case), shown in \Fig{IMFslope_vs_r}. This greater diversity in birth ISM pressures is likely a consequence of the fact that stellar feedback is burstier in \him{} due to the shallower high-mass slope of the IMF in high-pressure environments, potentially leading to a less uniform ISM and thus a broader range of ISM pressures.

\subsubsection{Mass-to-light excess radial gradients}
\label{sec:MLE_grads}

Many observational IMF studies parametrize the IMF via the excess $M/L$ ratio relative to that expected given a fixed reference IMF. We refer to this quantity as the $M/L$-excess, or MLE. The MLE is directly related to what dynamical IMF studies actually measure, and is easily computed given a best-fit IMF in spectroscopic studies.  We define the MLE in the $r$-band as
\be
\MLEkroupa{} = \frac{(M/L_r)}{(M/L_r)_{\rm Kroupa}}.
\label{eqn:MLEkroupa}
\ee
Note that here we use a non-logarithmic definition and compare $M/L$ relative to a Kroupa IMF (in contrast to the logarithmic, Salpeter-relative definition used in Papers I and II) to facilitate comparison with observational studies that tend to use the same definition.

In the upper panel of \Fig{MLE_vs_r} we plot \MLEkroupa{} measured within 2D projected elliptical concentric shells as a function of $r/r_e$ for our Sigma150 galaxies at $z=0.1$. For both simulations, the MLE is consistent with a Kroupa IMF at large radii (greater than a few $r_e$), gradually transitioning to larger values toward the centre. For \lom{}, all galaxies show clear negative radial gradients, while \him{} galaxies show much greater diversity: some have negative MLE gradients as strong as those in \lom{}, while others show no gradient at all. This diversity occurs despite the strong radial gradients in the IMF high-mass slope for the \him{} simulation, and is likely due to the additional dependence of the MLE on age when the high-mass slope is shallower than that for the reference IMF (Papers I and II). The lower panel of \Fig{age_pressure_vs_r} shows that the mean $r$-band light-weighted age exhibits strong variation within $r_e$ of the Sigma150 galaxies in \him{}, leading to the diversity in MLE. For both simulations, these radial trends in the MLE explain the aperture effects in the MLE-$\sigma_e$ relation we saw in Paper I. 

We compare our radial MLE trends with those for various observed galaxies in the upper panel of \Fig{MLE_vs_r}. As a yellow line we show the radial MLE gradient of M87 derived spectroscopically by \citet{Sarzi2018}, where we have performed a by-eye fit through their data points, assigning it 0.2 dex scatter. \citet{Oldham2018a} also obtain a negative gradient when inferring the spatially-resolved MLE in M87 dynamically (yellow points), which is offset systematically to lower MLE values by about 0.2 dex relative to the \citet{Sarzi2018} result. We also compare with spectroscopic results for XSG1 \citep[pink points;][]{LaBarbera2016}, NGC 1399 \citep[grey points;][]{Vaughan2018b}, NGC 4552 \citep[purple points;][]{Martin-Navarro2015b}, and NGC 1277 \citep[black points;][]{Martin-Navarro2015d}. For the latter two studies we convert $\Gamma_{\rm b}$ to \MLEkroupa{} assuming their published age gradients.  Finally, as a cyan line we show the radial MLE trend of \citet{vanDokkum2017}, which is a fit to the spectroscopically-determined MLE as a function of radius for 6 massive ETGs with $\sigma \sim 200-340\kms$.

For most of these observed systems, the MLE varies from super-Salpeter in the centre to Chabrier-like at around $0.4\,r_e$, while NGC 1399 and NGC 1277 remain Salpeter-like out to at least $\approx 1\,r_e$. Our simulated galaxies make qualitatively the same transition, but at larger radii, near or slightly above $1\,r_e$. Given the wide diversity in the observed trends, it is difficult to rule out either IMF parametrization with this test. Indeed, \lom{} galaxies rise to larger values of MLE than \him{}, which may be more consistent with most of these observed trends at the smallest radii. On the other hand, \him{} is more consistent with the Kroupa-like MLE values in some of the observations at $r/r_e \approx 1/3$ to 1, and with the shallow MLE gradient in NGC 4552 at all radii. The MLE values inferred for NGC 1399 are much larger than those for the other observed and simulated ETGs; the reason for this is unclear, but \citet{Vaughan2018b} speculate that these variances could be due to either differences in SPS modelling or the stochastic nature of galaxy formation. We thus conclude that both \lom{} and \him{} are consistent with the overall observed radial MLE gradients (although in Section \ref{sec:fdwarf_grads} we show that \him{} agrees with the MLE gradients derived from these spectroscopic studies for the wrong reasons). We reiterate that these radial variations were not considered in the calibration of our IMF variation prescriptions. 

Also of interest is the cumulative MLE measured within circular apertures of increasing radius, which we plot in the lower panel of \Fig{MLE_vs_r}. Again we see negative radial gradients, but with shallower slopes since the outer bins now contain the light from the central regions as well. We compare with spectroscopic results for six ETGs by \citet{vanDokkum2017} and for XSG1 from \citet{LaBarbera2016}. Both simulations agree with these observations at $r_e$, which is unsurprising given that they were calibrated to match the observed MLE$_r-\sigma_e$ relation (though only using data from \citealt{Cappellari2013b}), where MLE$_r$ is measured within $r_e$. At smaller radii, the simulated gradients are shallower than the observed trends. However, \lom{} galaxies are in excellent agreement with XSG1, and given the diversity in MLE gradients inferred from observations shown in the upper panel of \Fig{MLE_vs_r}, it is unclear if these discrepancies are robust.

Note as well that some studies find evidence for a lack of radial IMF gradients in high-mass ETGs. For example, \citet{Davis2017} use gas kinematics to infer the dynamical MLE radially within 7 ETGs, finding that, although the IMF seems to vary between galaxies, there is no systematic radial gradient. This data would thus support the \him{} model, in which it is possible to have flat MLE profiles even with significant gradients in the IMF slope, owing to the age-sensitivity of the relationship between MLE and high-mass IMF slope.

\subsubsection{Dwarf fraction radial gradients}
\label{sec:fdwarf_grads}

Another method of diagnosing the IMF is via the fraction of mass contributed by low-mass relative to high-mass stars.  This quantity is much closer to what is actually measured when inferring the IMF from gravity-sensitive stellar absorption features, compared with the IMF slope or MLE. Since in the old stellar populations present in typical ETGs, we expect that stars with $m \gtrsim 1\Msun$ should have died off, we define this fraction as
\be
F_{0.5,1} = \frac{ \int_{0.1}^{0.5}M\Phi(M)dM}{\int_{0.1}^{1}M\Phi(M)dM}.
\label{eqn:F051}
\ee
where $\Phi(M)$ is the IMF. We already showed in Paper I that this fraction, measured galaxy-wide, increases with $\sigma_e$ in \lom{} galaxies, in agreement with spectroscopic IMF studies \citep[e.g.][]{Conroy2012b, LaBarbera2013}. In the upper panel of \Fig{fdwarf_vs_r} we plot $F_{0.5,1}$ as a function of radius for our Sigma150 galaxies at $z=0.1$. Consistent with the IMF slope gradients seen in \Fig{IMFslope_vs_r}, we find a strong negative gradient in $F_{0.5,1}$ for \lom{}, transitioning from Salpeter to Chabrier-like at around $r_e$. For \him{}, the trend is much shallower because $F_{0.5,1}$ is  not very sensitive to high-mass slope variations. Interestingly, \him{} shows a shallow but positive radial $F_{0.5,1}$ gradient due to the shallower high-mass IMF slopes in the galaxy centres (see \Fig{IMFslope_vs_r}).

We compare these $F_{0.5,1}$ radial profiles with results for M87 \citep[yellow line;][]{Sarzi2018},  XSG1 \citep[pink points;][]{LaBarbera2016}, NGC 4552 \citep[purple points;][]{Martin-Navarro2015b}, and NGC 1277 \citep[black points;][]{Martin-Navarro2015d}, where for all of these studies we have converted the radial profiles of the high-mass IMF slope of a ``Bimodal'' IMF ($\Gamma_{\rm b}$) to profiles in $F_{0.5,1}$. All of these studies find negative gradients in $F_{0.5,1}$, consistent with \lom{}, although our simulated trend is shifted toward larger radii. \him{}, on the other hand, is inconsistent with these studies, implying that a prescription that varies only the high-mass slope of the IMF towards  values shallower than Salpeter is unable to explain these observations.

An alternative definition of the dwarf-to-giant ratio is the fraction of the {\it luminosity} that is contributed by low-mass ($m<0.5\Msun$) dwarf stars, $\fdwarf$. It has been shown by \citet{Alton2017} that this quantity correlates very strongly with the equivalent widths of IMF-sensitive stellar absorption features, and thus may be a better diagnostic of the low-mass regime of the IMF than $F_{0.5,1}$. However, since $\fdwarf$ by itself is also dependent on age, we normalize $\fdwarf$ for our galaxies by $\fdwarfSalp$, the value that would have been obtained had the stars evolved instead with a Salpeter IMF, given the same ages and metallicities. Our ($r$-band-weighted) $\fdwarf/\fdwarfSalp$ radial profiles are presented in the lower panel of \Fig{fdwarf_vs_r}, where we find qualitatively the same gradients as in the case of $F_{0.5,1}$.

\citet{Alton2018} spectroscopically measure the radially-resolved IMF in stacked spectra of 7 ETGs, where they parametrize the IMF by $\fdwarf$. We show their main results as brown squares in \Fig{fdwarf_vs_r}. They do not provide $\fdwarfSalp$ for these stacks, so we have normalized their results by the $\fdwarfSalp$ at an age of 10 Gyr presented in Table C2 of \citet{Alton2017}, multiplied by the ratio age/(10 Gyr). This procedure assumes a linear relationship between $\fdwarfSalp$ and age, which is the case for our variable IMF simulations. The error bars include the uncertainty in the age. 

Owing to the large scatter of the \citet{Alton2018} results, they are consistent with both a flat IMF gradient and the steep negative $\fdwarf$ gradient in \lom{} galaxies. This apparent paradox is also due in part to the fact that for \lom{}, $\fdwarf$ transitions to a Kroupa value at around $r_e$, which is where the observations stop due to limited $S/N$. Deeper observations toward the outskirts of these galaxies would be required to establish if their IMF gradients are truly flat.

Note as well that \citet{Alton2018} find that $\fdwarf$ is sensitive to assumptions made in the modelling of these galaxies. To demonstrate this sensitivity, as grey points we show the results of \citet{Alton2018} where they impose flat priors on $\fdwarf$ itself (rather than on the low-mass IMF slopes; brown diamonds). This procedure increases the $\fdwarf/\fdwarfSalp$ values in the central regions. For completeness, we also include the results of \citet{Alton2017} as brown circles, where the same analysis was performed on the same galaxies, except with fewer absorption features and a less up-to-date SPS model. Here $\fdwarf$ becomes super-Salpeter at small radii. Thus, spectroscopic inferences of the IMF are quite sensitive to the methods used (even on the same galaxies), leading to potentially strong systematic errors. A better understanding of the systematics involved in spectroscopic modelling, as well as higher resolution in simulations, will be required to quantitatively compare the IMF in the inner regions of simulated and observed galaxies.

Overall, we find that radial IMF gradients are a natural prediction of models in which the IMF is a function of local physical conditions in the ISM (in our case the pressure at which the stellar populations are born). For low-mass slope variations, our radial gradients in the IMF low-mass slope, MLE, and dwarf-to-giant ratio are all in agreement with observational IMF studies which find such gradients, but this model may not be able to explain studies that do not find such gradients. \him{}, on the other hand, is perhaps more consistent with the diversity in the MLE radial gradients from observational studies which can range from strongly negative to flat, but is inconsistent with the negative radial dwarf-to-giant ratio gradients inferred from spectroscopic studies. Stronger consensus from observational studies will be pivotal in further constraining proposed IMF variation prescriptions.

\subsection{Radial gradients in stellar population properties}
\label{sec:starprops_gradients}

% fig 9 - abundances vs r
\begin{figure}
   \centering
  \includegraphics[width=0.49\textwidth]{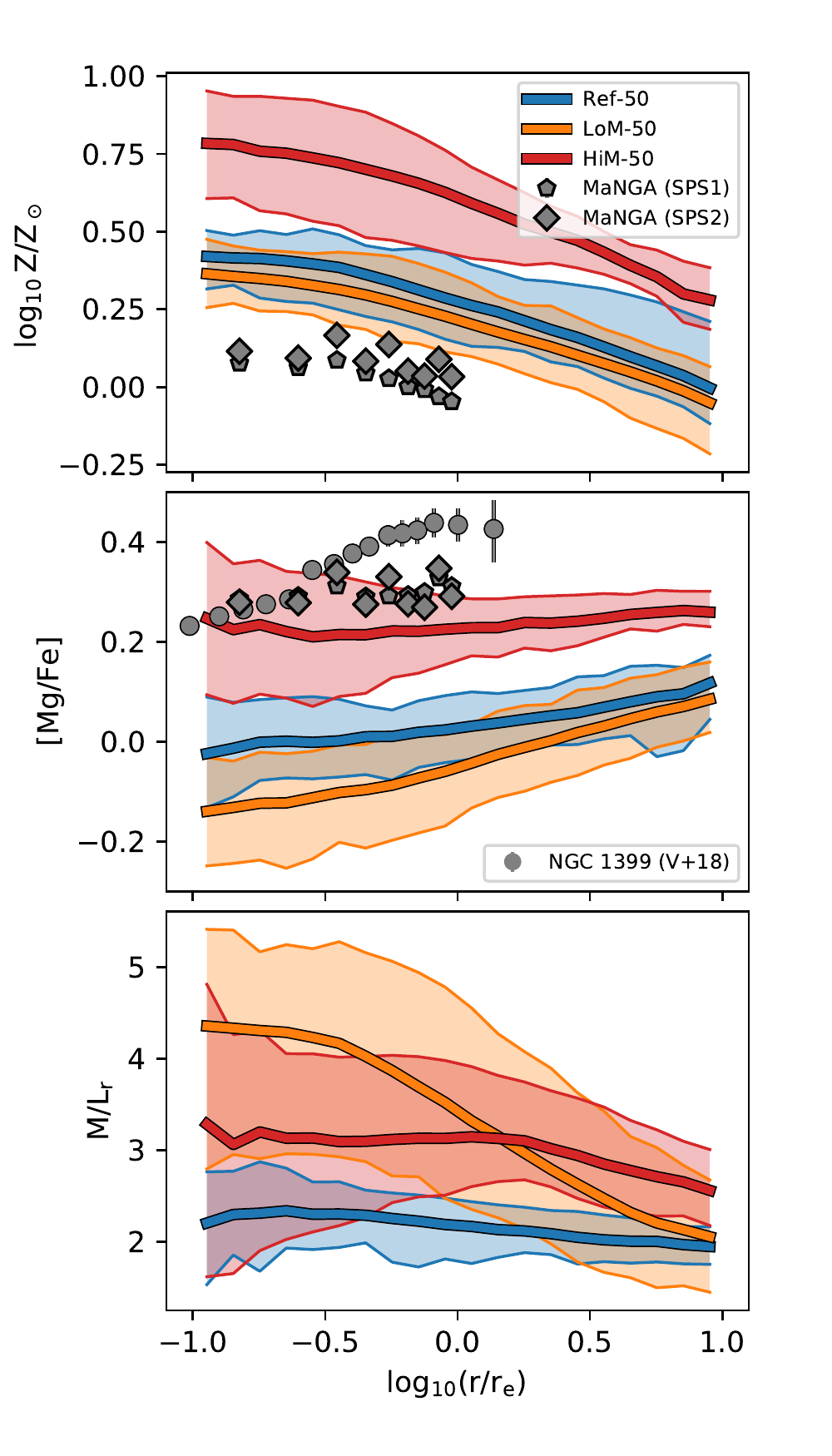}
 \caption{As \Fig{age_pressure_vs_r} but showing, from top to bottom, radial gradients in the stellar metallicity, stellar $\alpha$-enhancement [Mg/Fe], and $r$-band stellar $M/L$ ratio for our Sigma150 ETGs. We show metallicity and [Mg/Fe] gradients for high-mass galaxies from SDSSIV-MaNGA from \citet{Parikh2018}, where pentagons and diamonds correspond to using two different SPS models. The [Mg/Fe] radial gradient in NGC 1399 is shown as grey circles \citep{Vaughan2018b}.  While negative metallicity gradients are preserved in all of our simulations, the [Mg/Fe] gradients are more positive and nearly flat for \lom{} and \him{}, respectively. $M/L$ gradients are negative in \lom{}, while in \him{} the gradients are much more diverse within $r_e$. }
  \label{fig:abundances_vs_r}
 \end{figure}

Since the IMF varies strongly with radius, it is also interesting to investigate how the gradients of other stellar properties are affected by variations in the IMF. Here we investigate gradients in metallicity, [Mg/Fe], and $r$-band $M/L$ ratio for each galaxy binned radially in the same logarithmically-spaced elliptical bins as in the previous section. We plot these results in \Fig{abundances_vs_r} for our Sigma150 samples in Ref-50, \lom{}, and \him{} at $z=0.1$.

The upper panel of \Fig{abundances_vs_r} shows the radial metallicity profiles of our simulated galaxies, all of which have strongly negative gradients. Since, as we saw in Paper I, \lom{} galaxies showed the same stellar mass$-$metallicity relation as Ref-50, it is unsurprising that they also have similar metallicity gradients. While \him{} galaxies exhibit the same gradient, the profiles are normalized to larger metallicity by $\approx 0.2$ dex at all radii. This is the result of a higher production of metals due to a top-heavy IMF. Indeed, the offset appears to be strongest at small radii, where the IMF is most top-heavy in \him{}. For comparison we show the metallicity gradients in high-mass ETGs found with SDSSIV-MaNGA by \citet{Parikh2018} (symbols as in \Fig{IMFslope_vs_r}). The shapes of the simulated profiles agree well with the observed results, but with some systematic offsets in normalization. However, these offsets are consistent with the factor two uncertainty in the nucleosynthetic yields in the simulations \citep{Wiersma2009b} as well as the systematic offsets between different metallicity calibrators which can be as high as 0.7 dex \citep{Kewley2008}, and are thus not particularly constraining.

In the middle panel of \Fig{abundances_vs_r} we show radial [Mg/Fe] profiles for our simulated ETGs. The profile of Ref-50 is fairly flat out to $1\,r_e$, and slowly rises beyond it. That of \him{} is even flatter but, as for the metallicity gradients, it is offset by $\approx 0.2-0.3$ dex. \lom{}, however, shows steeper positive radial gradients due to a deficit of $\alpha$-elements in the inner regions. These results are in broad agreement with [Mg/Fe] gradients in observed galaxies which seem to vary on a case-by-case basis from either being flat \citep{Mehlert2003, Parikh2018}, to weakly positive \citep{Spolaor2008, Brough2007}, or strongly positive \citep[as is the case for NGC 1399;][]{Vaughan2018b}. The offsets between our simulations relative to Ref-50 are consistent with those for the (galaxy-wide) [Mg/Fe]$-\sigma_e$ relation seen in Paper I, and are due to the decreased (increased) number of type II SNe resulting from a steeper (shallower) IMF slope in \lom{} (\him{}). 

The $r$-band $M/L$ ratio radial profiles are shown in the lower panel of \Fig{abundances_vs_r}. While in Ref-50 the gradient is generally flat with $M/L_r \approx 2-3$ $\Msun/\Lsun$ at all radii, in \lom{} it increases by a factor $\approx 2$ toward the centre, as expected given the MLE gradients seen in \Fig{MLE_vs_r}. \him{} ETGs have a much greater diversity in $M/L$, ranging from 1 to 5 below $0.3\,r_e$. These non-constant $M/L$ ratios may have important consequences for dynamical mass measurements \citep[see][]{Bernardi2018b, Sonnenfeld2018, Oldham2018b}, particularly when such masses are used to infer the IMF \citep{Cappellari2013b}. Here we have shown that in the case of IMF variations, the assumption of a constant $M/L$ within $r_e$ is not justified.

% fig 10 - IMF vs abundances
\subsection{IMF vs local quantities}
\label{sec:IMF_vs_local_starprops}
  \begin{figure*}
   \centering
  \includegraphics[width=0.45\textwidth]{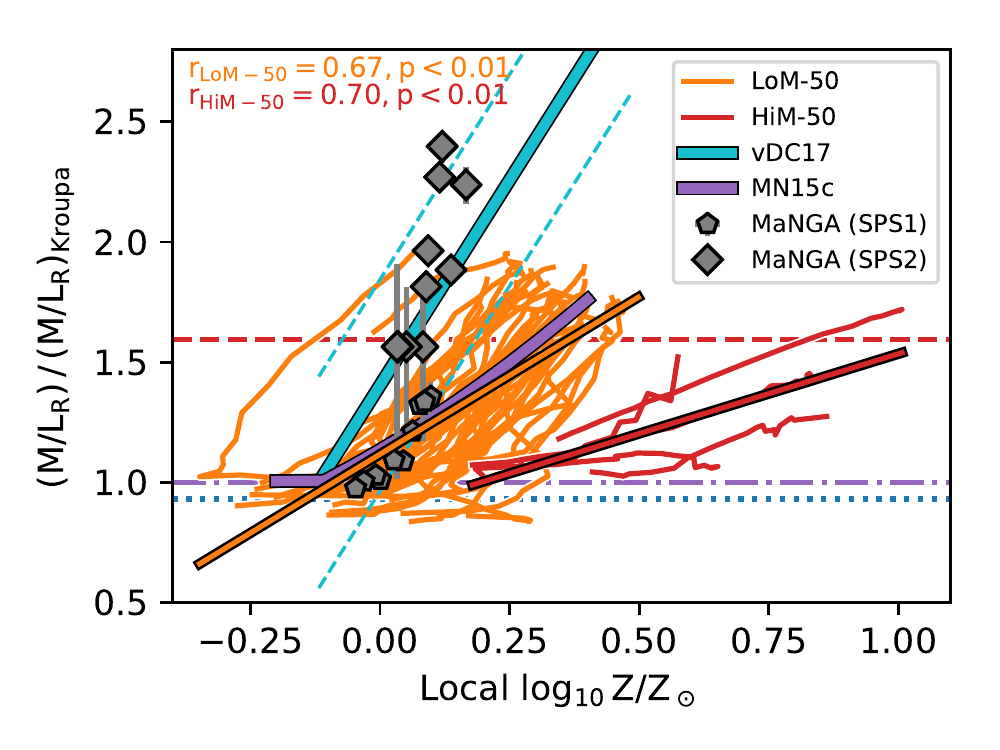}
  \includegraphics[width=0.45\textwidth]{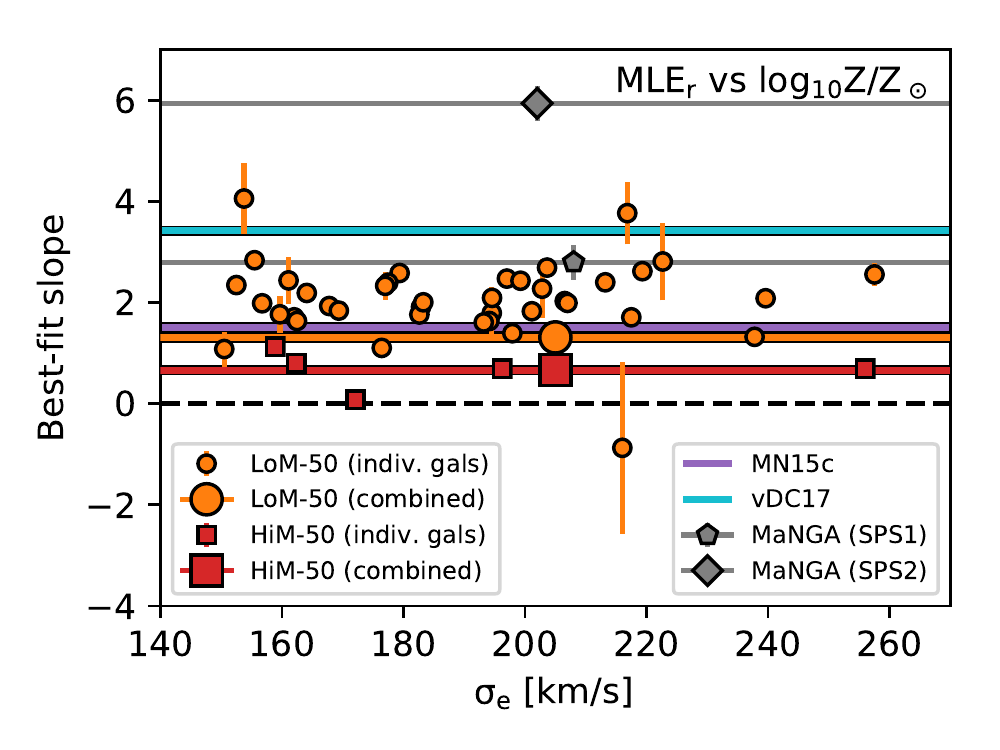}
   \includegraphics[width=0.45\textwidth]{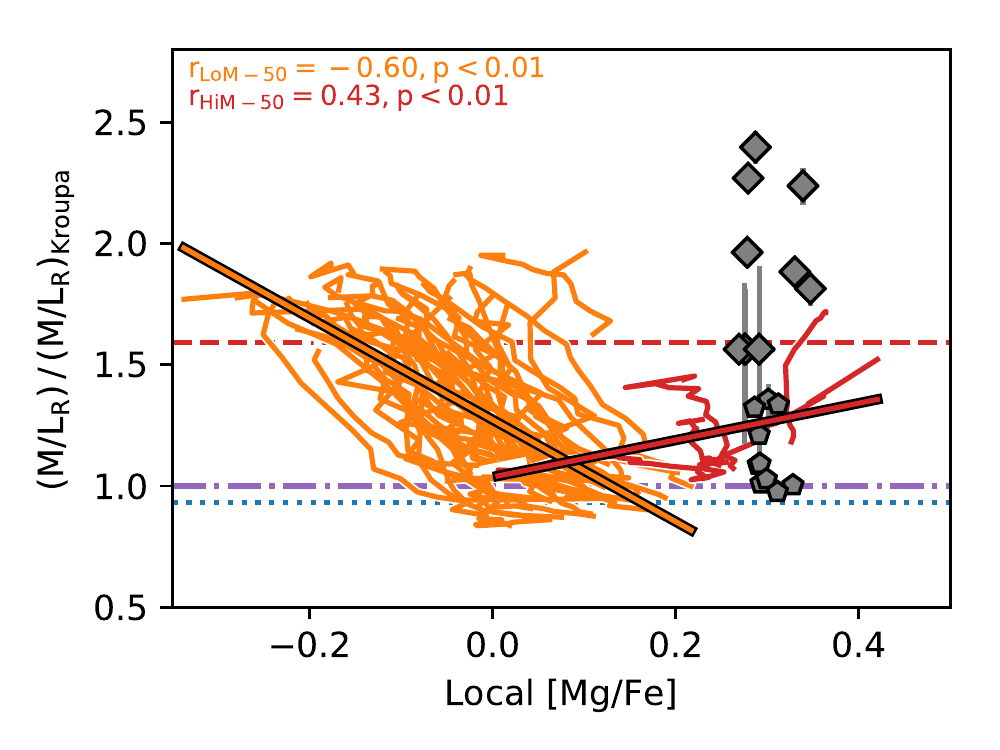}
  \includegraphics[width=0.45\textwidth]{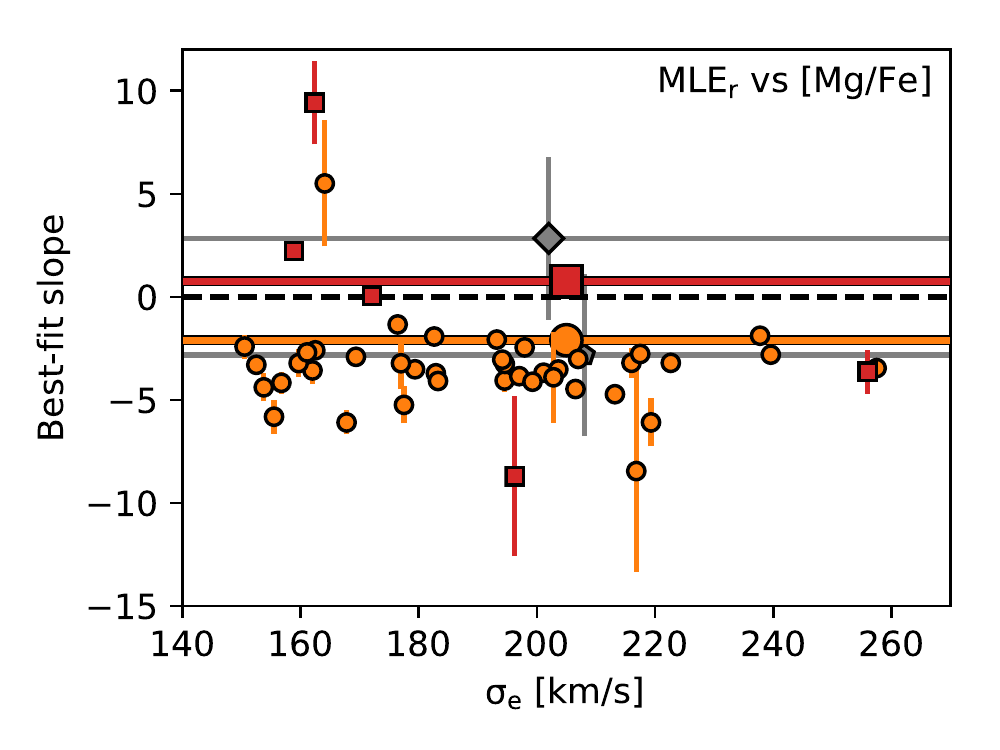}
  \includegraphics[width=0.45\textwidth]{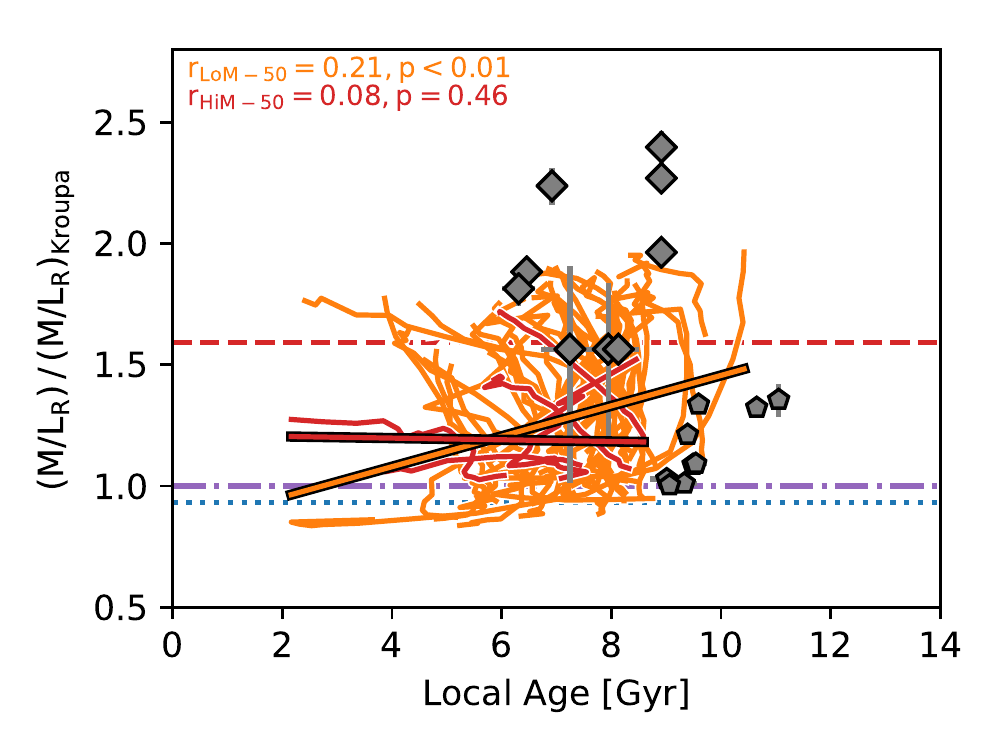}
  \includegraphics[width=0.45\textwidth]{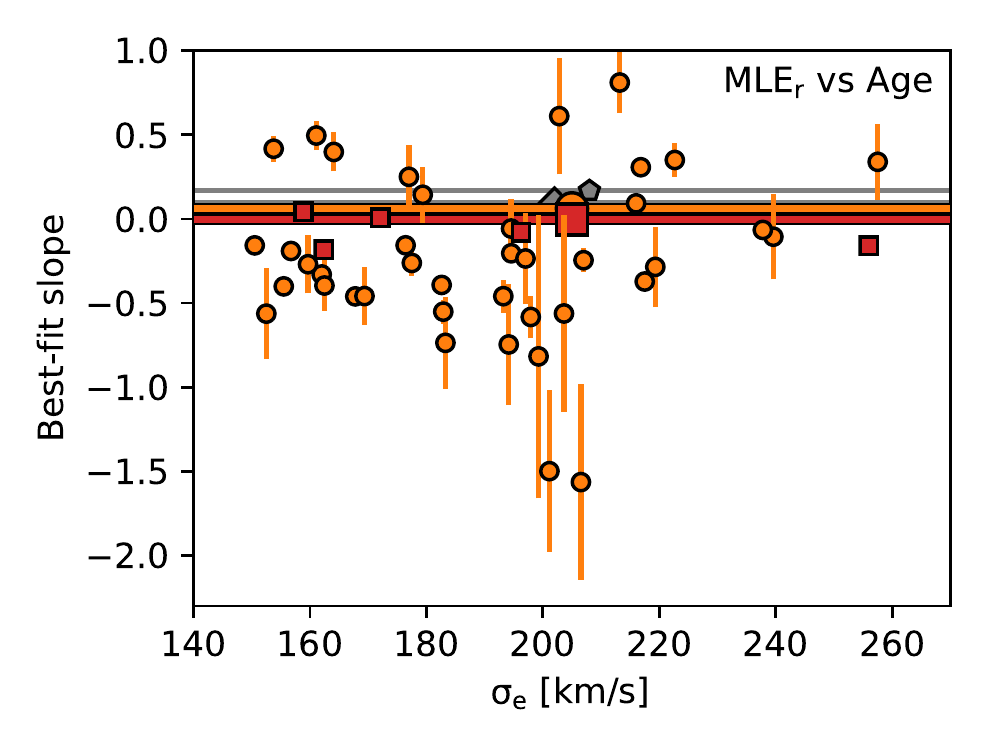}
 \caption{{\bf Left column:} Local excess $r$-band $M/L$ relative to a Kroupa IMF (\MLEkroupa{}) as a function of local stellar metallicity (upper row), stellar [Mg/Fe] (middle row), and stellar age (lower row) within ETGs with $\sigma_e>150\kms$ from \lom{} (orange) and \him{} (red) at $z=0.1$. All quantities are $r$-band light-weighted. Each thin coloured line shows values for an individual galaxy in logarithmically-spaced radial bins from 1 kpc to $10\,r_e$. Least absolute deviation (LAD) fits to all galaxies simultaneously are shown thicker, outlined in black, with Spearman $r$-values and their $p$-values indicated in each panel. Trends for high-mass ETGs from SDSSIV-MaNGA for two different SPS models from \citet{Parikh2018} are shown as grey pentagons and diamonds, respectively. The observed local MLE-metallicity relation found by \citet{Martin-Navarro2015c} is shown as a purple solid line and that for 6 massive ETGs from \citet{vanDokkum2017} is shown as a solid cyan line with intrinsic scatter shown as cyan dashed lines. Expected MLE values for fixed Salpeter, Kroupa, and Chabrier IMFs are indicated with horizontal red-dashed, purple-dot-dashed, and blue-dotted lines, respectively.  {\bf Right column:} Logarithmic slopes of LAD fits to the local \MLEkroupa{} as a function of local properties within individual galaxies with $\sigma_e > 150\kms$ in \lom{} (orange circles) and \him{} (red squares), plotted as a function of $\sigma_e$. Each small point in the right column corresponds to a thin line in the left, while larger symbols show the slopes of the LAD fits to all of these galaxies simultaneously. \MLEkroupa{} systematically correlates positively with local metallicity in both simulations, in qualitative agreement with the observations. The correlation with local [Mg/Fe] is negative and positive in \lom{} and \him{}, respectively, but for \him{} it differs substantially between galaxies. When combining results from all galaxies, \MLEkroupa{} increases weakly with local age for \lom{}, but on average decreases for individual galaxies. No correlation with local age is found for \him{} galaxies.}
  \label{fig:IMF_vs_abundances}
 \end{figure*}

We now study the correlation between the IMF (parametrized by the MLE), and local galaxy properties. To this end we investigate the trends between MLE and metallicity, [Mg/Fe], and age, measured in the same logarithmically-spaced elliptical annuli as in the previous subsections. Crucially, the question of whether or not such correlations exist can depend on how the data are combined. First we ask, if we consider the data from all of the radial bins in all of the Sigma150 galaxies simultaneously, is there a correlation between MLE and local properties? This procedure is similar to what is done by some observational studies such as \citet{Martin-Navarro2015c} or \citet{vanDokkum2017}.

In the top-left panel of \Fig{IMF_vs_abundances} we show \MLEkroupa{} as a function of stellar metallicity. Here each thin line shows the trend for an individual galaxy in radial bins, while thick solid lines show the best least absolute deviation (LAD) fit result when fitting to all galaxies simultaneously, with corresponding Spearman $r$-values and their $p$-values indicated in the upper left of each panel for \lom{} (orange) and \him{} (red). For both simulations, we see overall a significant positive relation, likely due to the fact that both the IMF and metallicity vary radially monotonically. For comparison, we show the same trend for the 6 ETGs studied by \citet{vanDokkum2017} as a thick cyan line\footnote{We have shifted the \citet{vanDokkum2017} values to the right by 0.3 dex to convert from [Fe/H] to $Z/Z_\odot$, given their fairly consistent [Mg/Fe] values of 0.3 dex and the relation from \citet{Thomas2003}.}, as well as the relation by \citet{Martin-Navarro2015c}, and that for the highest-mass bin ($10.5 < \log_{10}\Mstar/\Msun < 10.8$) of SDSS-MaNGA by \citet{Parikh2018}. All of these studies find strong positive correlations of the IMF slope with increasing local metallicity, in good agreement with our findings. However, the shallower slope of this relation for \him{} galaxies is in slight tension with the observations. Note that our IMF (and MLE) does not depend physically on metallicity. Thus, a strong, tight correlation with metallicity can be expected even if the IMF varies with another property of the ISM, in our case pressure.

The above analysis may be sensitive to correlations between the MLE and global galaxy properties if the range of global values among galaxies is larger than the range of local values within them. To eliminate such effects, we also measure the slope of this correlation for each individual galaxy. We show these results in the upper right panel of \Fig{IMF_vs_abundances}. For nearly every Sigma150 ETG in both \lom{} and \him{}, we find a significant positive correlation between the local MLE and local metallicity. We compare with the slopes of the observed relations shown in the left panel. For \citet{Parikh2018} we manually perform a least-squares fit to their data only for the highest-mass bin, rather than over all 3 mass bins as in that paper, as the latter analysis would be sensitive to a global correlation between MLE and age (as can be seen by the lack of overlap in radially-resolved age in their highest and lowest mass galaxy bins in their Fig. 17). The slopes of our relations for individual galaxies in \lom{} are consistent with those observed, while the slopes for \him{} galaxies, albeit  consistently positive, are shallower than the observed trends. Note as well that, strictly speaking, it is not completely fair to compare \him{} with these observational studies since they measure the IMF spectroscopically, constraining the dwarf-to-giant ratio. Indeed, the dwarf-to-giant ratio in \him{}, being relatively constant with radius (\Fig{fdwarf_vs_r}) would not correlate with changes in local stellar properties. We also do not see any correlation of these best-fit slopes with $\sigma_e$ for these high-$\sigma_e$ galaxies.

We also perform the same investigation into the correlation between local \MLEkroupa{} and local [Mg/Fe], shown in the middle row of \Fig{IMF_vs_abundances}. Here we see strikingly different behaviour. When taking all of the galaxies together (solid thick lines in the middle left panel), we find a strong negative correlation of \MLEkroupa{} with [Mg/Fe] in \lom{}, while that for \him{} galaxies is positive. This difference between the simulations opens up a novel method of discriminating between these IMF parametrizations.  When looking at galaxies individually, this negative relation persists for the majority of \lom{} galaxies, but no systematic trend is obvious for our 5 \him{} ETGs. Thus, the positive relation seen when combining results from these five \him{} galaxies likely reflects the strong global MLE--[Mg/Fe] relation for \him{} ETGs (see Fig. 2 of Paper II). Care should thus be taken when inferring local IMF relations to first remove global trends between galaxies.

These correlations between the MLE and [Mg/Fe] may be in tension with radially-resolved IMF studies, in which typically no significant correlation between the IMF and local [Mg/Fe] is found (e.g. \citealt{Martin-Navarro2015c, vanDokkum2017, Parikh2018}, but see \citealt{Sarzi2018}). However, the best-fit slopes from \citet{Parikh2018}, while consistent with zero, are also consistent with our findings, both for \lom{} and \him{} (middle right panel). Thus, the correlations between the local MLE and local [Mg/Fe] may be washed out in some current studies by observational uncertainties.
 
In the lower row of \Fig{IMF_vs_abundances} we perform the same analysis but for the correlation between local \MLEkroupa{} and local stellar age. For \lom{}, when considering the entire sample together, we find a weak positive correlation with age (shown by the solid orange line and corresponding Spearman $r$-value in the lower left panel). However, the same is not true for individual galaxies (orange points in the lower right panel), where the best-fit slope tends to scatter around 0 at all $\sigma_e$. For \him{} we do not find any systematic correlation with local age in either case. Thus, for both simulations we see no significant systematic correlation between MLE and age within individual galaxies. Indeed, the positive trend for \lom{} seen in the lower left panel merely reflect the strong dependence of the MLE on age when measured ``galaxy-wide'' (see Fig. 2 of Paper II). These differences are subtle, but are extremely important in  trends between the IMF and local properties in observational studies, especially when combining results from galaxies with a wide range of global properties, as is often necessary to obtain sufficiently high $S/N$ ratio out to large radii.

Overall, these results highlight the importance of removing global trends between the MLE and integrated galaxy properties before interpretting trends within galaxies. For both simulations we find strong positive correlations of MLE with local metallicity, and the average negative (for \lom{}) and positive (for \him{}) correlations with [Mg/Fe] will be useful in discriminating between scenarios in which the IMF becomes either bottom-heavy or top-heavy in high-pressure environments.

\section{Redshift dependence of galaxy properties}
\label{sec:redshift}

In Paper I we investigated the galaxy-wide correlations between IMF-related diagnostics such as the MLE, $F_{0.5,1}$, and ionizing flux with global galaxy properties such as $\sigma_e$ and star formation rate (SFR), finding good agreement with observations. In this section we switch from the spatially-resolved properties discussed in the previous sections to investigate the evolution of these (global) IMF scaling relations (\Sec{redshift_IMF}), as well as the effect that our IMF variation models have on the evolution of the cosmic properties in the simulations (\Sec{redshift_cosmic}).  

\subsection{ Redshift dependence of the IMF }
\label{sec:redshift_IMF}

%fig 11: redshift dependence of the MLE-sigma relation
 \begin{figure}
   \centering
\includegraphics[width=0.5\textwidth]{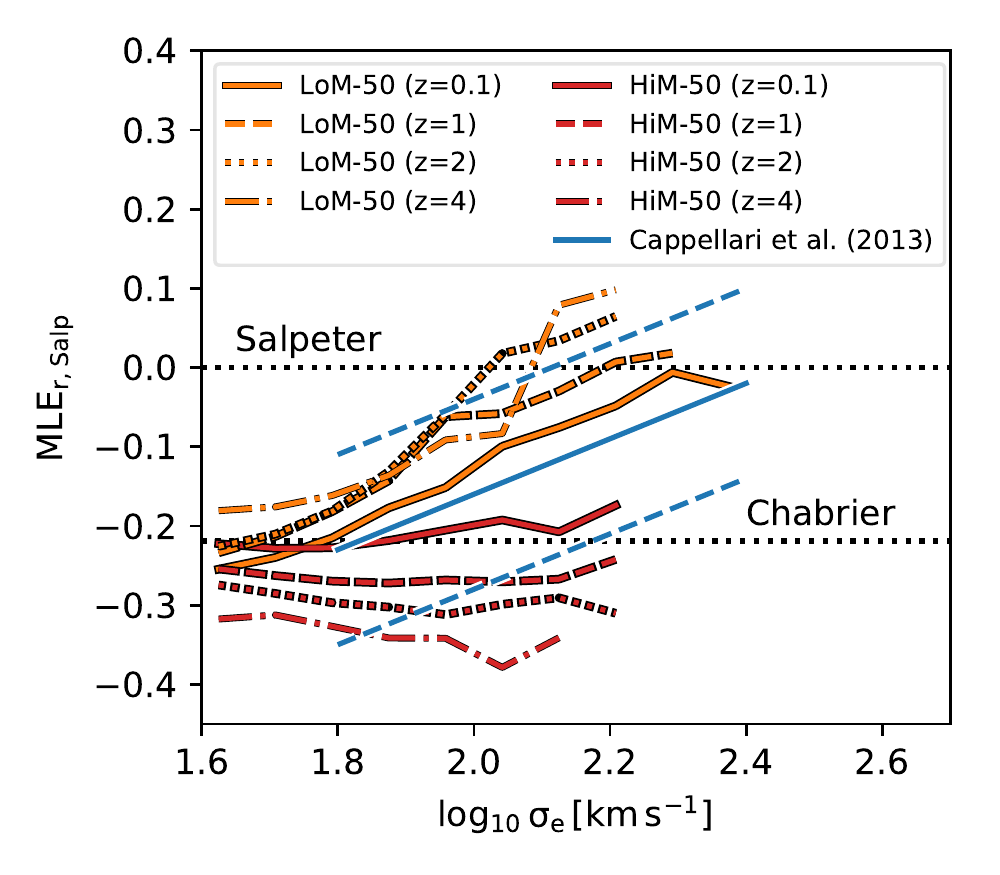}
 \caption{Galaxy-wide \MLEsalp{} as a function of $\sigma_e$ for \lom{} (orange) and \him{} (red) at $z=0.1$ (solid lines), $z=1$ (dashed lines), $z=2$ (dotted lines), and $z=4$ (dash-dotted lines) for all galaxies with $\sigma_e > 10^{1.6}\kms$ at each redshift. Thick lines shown medians for bins with more than 5 galaxies. All quantities are $r$-band light-weighted and measured within the 2D projected $r$-band half-light radius of each galaxy. A blue solid line indicates the \citet{Cappellari2013b} relation at $z\approx 0$, with dashed blue lines denoting intrinsic scatter. Horizontal black dotted lines mark the Salpeter and Chabrier $\MLEsalp{}$ values. For the \lom{} (\him{}) run, galaxies have a heavier (lighter) \MLEsalp{} with increasing redshift. }
  \label{fig:MLE_vs_z}
 \end{figure}

%fig 12: redshift dependence of the MLE-sigma relation - star histograms
 \begin{figure*}
   \centering
\includegraphics[width=0.45\textwidth]{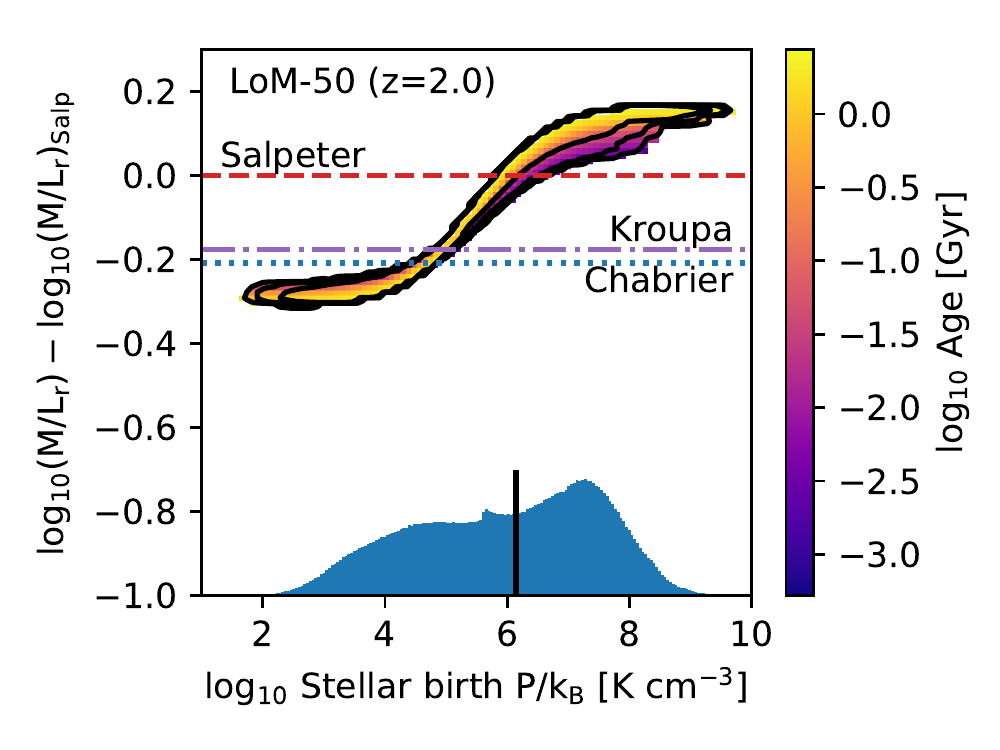}
\includegraphics[width=0.45\textwidth]{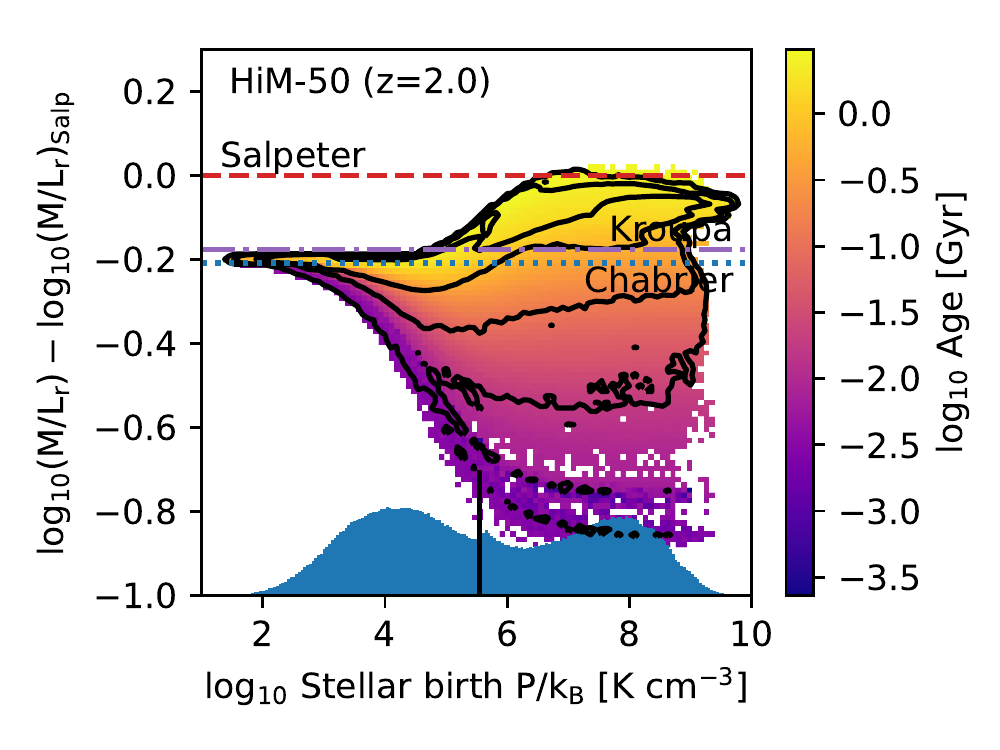}
\includegraphics[width=0.45\textwidth]{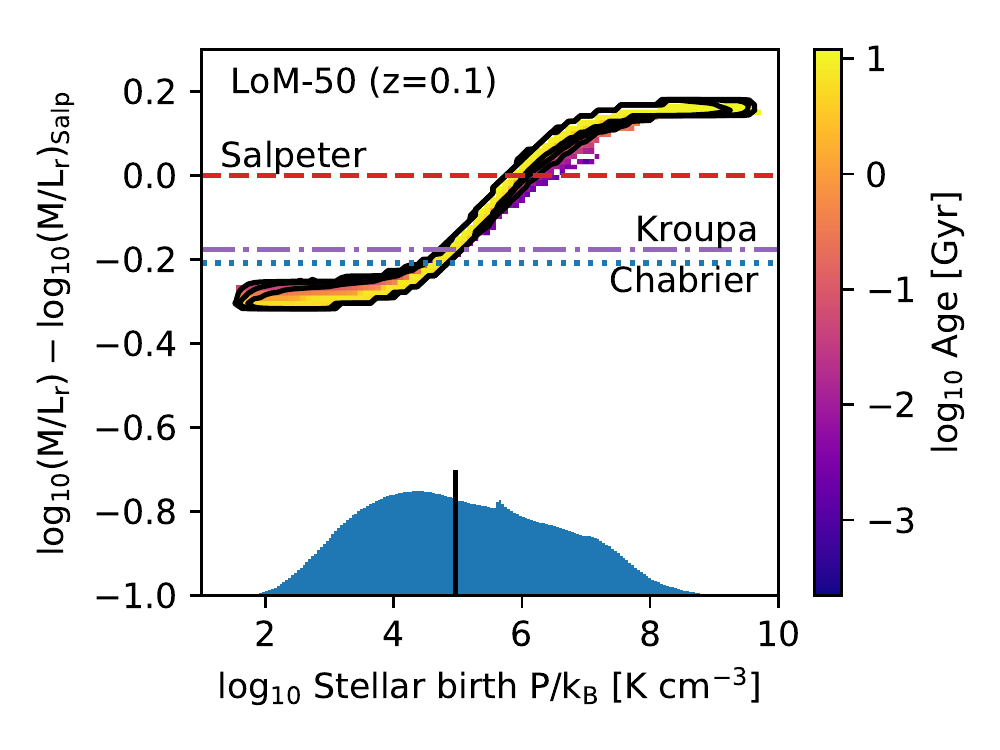}
\includegraphics[width=0.45\textwidth]{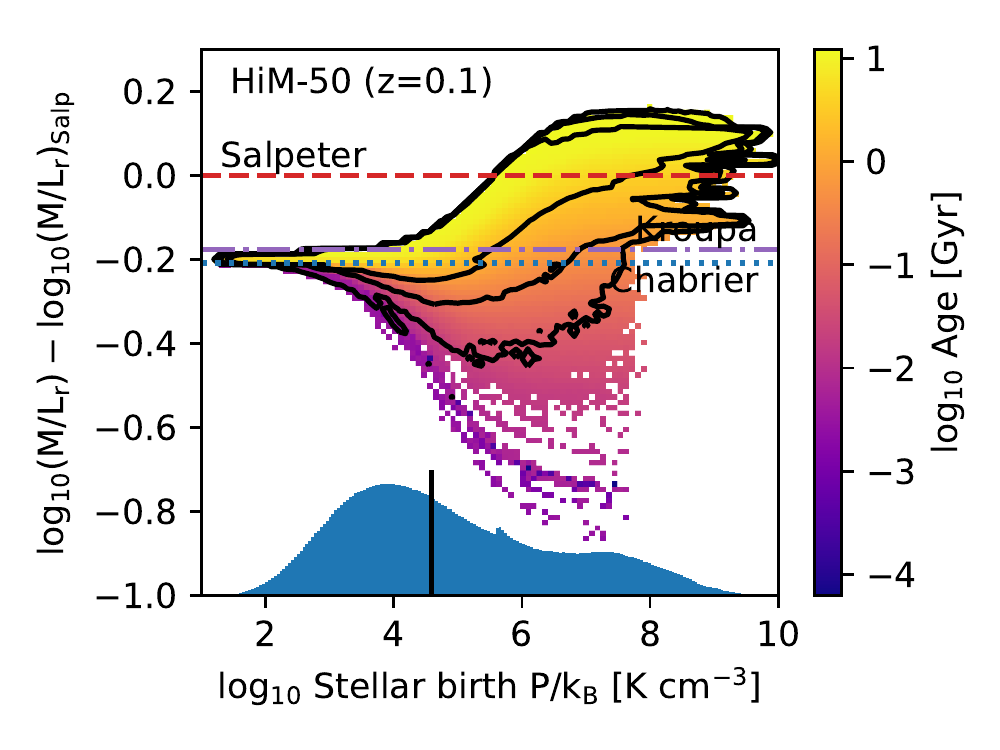}
 \caption{Redshift dependence of the \MLEsalp{}$-\sigma_e$ relation for individual stellar particles in the \lom{} (left column) and \him{} (right column) simulations. 2D histograms in the upper and lower rows show \MLEsalp{} as a function of birth ISM pressure for stars within the half-light radius of galaxies with $\sigma_e > 100\kms$ at $z=2$ and $z=0.1$, respectively. 2D bins are coloured by the mean age of stars within each bin. Contours show boundaries of (1000, 100, 10) stars per bin. Blue 1D histograms show the normalized distribution of birth ISM pressures on a linear scale; black horizontal lines show the medians of these distributions. In the \lom{} case, higher birth ISM pressures at high $z$ result in a larger fraction of dwarf stars than at $z=0.1$. As \MLEsalp{} is relatively independent of age for this IMF, the effect is immediate. In the HiM case, most stars born at high pressure have not lived long enough for the \MLEsalp{} to become ``heavy'' (i.e. for massive stars to die off, reducing the light and increasing the mass due to BHs/NSs), so they still have low $M/L$ ratios relative to Salpeter at this high redshift.}
  \label{fig:MLE_vs_z_particles}
 \end{figure*}

%fig 13: redshift dependence of the IMF-sigma relation
 \begin{figure}
   \centering
\includegraphics[width=0.5\textwidth]{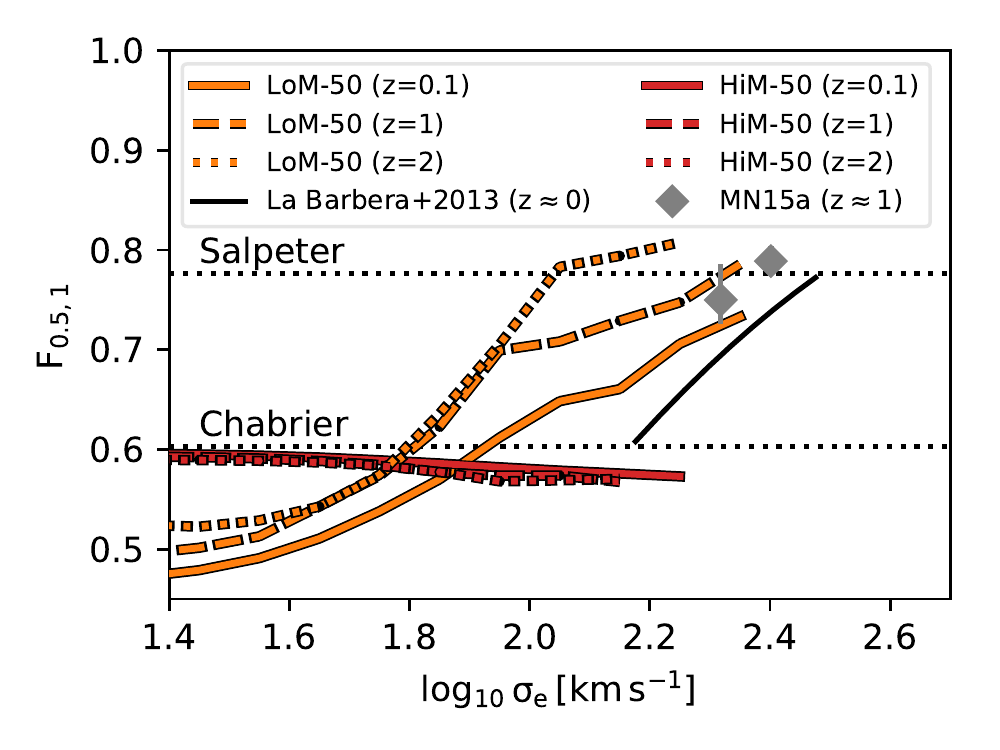}
 \caption{Redshift dependence of dwarf fractions in \lom{} and \him{}. The mass fraction of stars with $m<0.5\Msun$ relative to those with $m<1\Msun$ in the IMF, $F_{0.5,1}$, is shown as a function of $\sigma_e$. Line styles are as in \Fig{MLE_vs_z}. The result of \citet{LaBarbera2013} at $z\approx 0$ (assuming a bimodal IMF) is shown as a solid black line. Values for quiescent galaxies at $z\approx 1$ from \citet{Martin-Navarro2015a} are shown as grey diamonds with 1$\sigma$ error bars.  The $F_{0.5,1}-\sigma_e$ relation evolves significantly for \lom{}, but not for \him{}. }
  \label{fig:F05_vs_z}
 \end{figure}

% fig 14 - star-forming properties at high z
 \begin{figure}
   \centering
\includegraphics[width=0.5\textwidth]{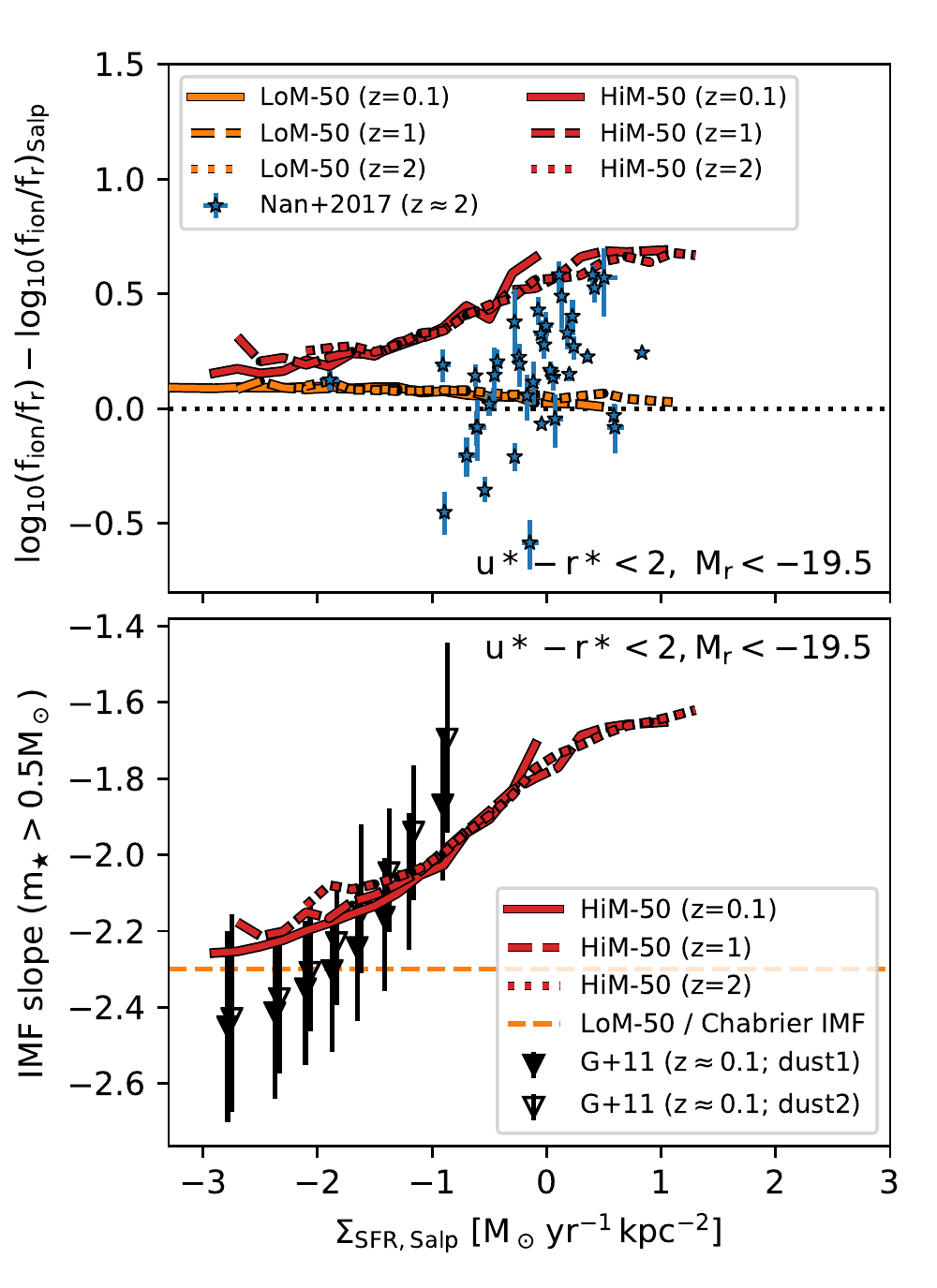}
 \caption{Upper panel: Ionizing flux relative to SDSS $r$-band flux, normalized by the value expected for a Salpeter IMF, as a function of Salpeter-reinterpreted star formation rate (SFR) surface density, $\Sigma_{\rm SFR,Salp}$, in bright ($M_r<-19.5$), star-forming (intrinsic $u^*-r^*<2$) galaxies in \lom{} (orange) and \him{} (red) at different redshifts.  Line styles are as in \Fig{MLE_vs_z}.  For reference we include observations of H$\alpha$ equivalent width relative to that expected for a Salpeter IMF for star-forming galaxies at $z\approx2$ by \citet{Nanayakkara2017}.
 Lower panel: FUV-weighted high-mass ($m>0.5\Msun$) IMF slope as a function of $\Sigma_{\rm SFR,Salp}$ for the same galaxy sample as in the upper panel.The high-mass slope for all \lom{} galaxies (the Chabrier value) is shown as a dashed orange line. For reference we include observations of local star-forming galaxies by \citet{Gunawardhana2011} for two different dust extinction models. The ionizing-to-$r$ band flux ratio and IMF slopes do not evolve appreciably with redshift at fixed $\Sigma_{\rm SFR,Salp}$ for either simulation, but for \him{} the relation extends to larger $\Sigma_{\rm SFR,Salp}$, and thus larger $f_{\rm ion}/f_{\rm r}$ and shallower high-mass IMF slopes, at higher redshift. The observed trends of shallower IMF slope or stronger H$\alpha$ EWs with increasing $\Sigma_{\rm SFR,Salp}$ at all redshifts are qualitatively consistent with our \him{} simulation.}
\label{fig:highmassslope_and_HalphaEW_vs_z}
 \end{figure}

Some models of IMF variation in the literature invoke time-dependent IMF variations to explain the enhanced dwarf-to-giant ratios as well as high metal enrichment in high-mass ETGs \citep[e.g.][first a top-heavy starburst, then a prolonged bottom-heavy mode]{Arnaud1992, Weidner2013, Ferreras2015, Martin-Navarro2016}. Since our IMF depends on pressure, and typical birth ISM pressures decrease with time, our average IMF is also implicitly time-dependent. Indeed, for our HiM IMF prescription, high-pressure starbursts will naturally be given a top-heavy IMF while lower-pressure, less rapid star formation will proceed with an IMF closer to Kroupa (bottom heavy).

We now investigate the evolution of IMF-related diagnostics in our simulations for unresolved, galaxy-averaged properties. As in Papers I and II, we compute these properties as $r$-band light-weighted mean quantities measured within a circular projected aperture of radius $r_e$. Also for consistency with Papers I and II, we redefine the MLE as
\be
\MLEsalp = \log_{10}(M/L_r) - \log_{10}(M/L_r)_{\rm Salp},
\label{eqn:MLEsalp}
\ee
which is effectively a rescaled version of \MLEkroupa{}.

In \Fig{MLE_vs_z}, we show the median relation between $\MLEsalp$ and $\sigma_e$ at $z=0.1$, 1, 2, and 4 (different line styles) for \lom{} (orange) and \him{} (red) for all galaxies with $\sigma_e>10^{1.6}\kms$ at each redshift. Despite being calibrated to roughly the same value at $z=0.1$, the redshift evolution of the $\MLEsalp-\sigma_e$ relation differs greatly between the two IMF prescriptions. While the MLE of high-$\sigma_e$ galaxies increases with redshift for \lom{} galaxies, it decreases for \him{} galaxies. At $z=2$, for example, the typical $\MLEsalp$ in \lom{} is $\approx 0.1$ dex higher than the $z=0.1$ relation, with most galaxies in this mass range having a super-Salpeter IMF. On the other hand, galaxies in \him{} are $\approx 0.1$ dex {\it lower} at $z=2$ than at $z=0.1$.

We can understand these differences via \Fig{MLE_vs_z_particles}, where we show $\MLEsalp$ as a function of stellar birth ISM pressure at $z=2$ and $0.1$, respectively, for stars within the half-light radius of galaxies with $\sigma_e > 100\kms$. Blue histograms show the distribution of birth ISM pressures for these stars. For the \lom{} simulation shown in the left column, we see that the $\MLEsalp-$pressure relation is nearly identical at high and low redshift, due to the age-independence of the MLE for this form of the IMF (see Fig. 5 of Paper I). However, the blue histograms show that birth ISM pressures of stars were typically much higher at $z=2$ than at $z=0.1$. This result is due to the fact that densities are in general higher at higher redshift, and thus the pressures at which stars form are also higher.

On the other hand, for the \him{} simulation, not only were the stars formed at higher pressure at high redshift, but also the shape of the $\MLEsalp-$pressure relation changes over time, becoming more ``heavy'' as time goes on. This effect is due to the age dependence of the MLE for an IMF with a shallow high-mass slope. Stellar particles at high pressure are born ``light'', with low MLE due to a prevalence of high-mass stars, but over time become ``heavy'' due to stellar evolution removing these bright stars and leaving behind stellar-mass BHs and NSs.

\citet{Martin-Navarro2015a} recently inferred from spectroscopic observations that the IMF of ETGs at $z\approx 1$ is consistent with that of ETGs at low redshift. To compare with their results, we plot in \Fig{F05_vs_z} the mass ratio between stars with $m<0.5\Msun$ and $m<1\Msun$ in the galaxy-averaged IMF, $F_{0.5,1}$ (Equation \ref{eqn:F051}), as a function of $\sigma_e$. As with the MLE, $F_{0.5,1}$ becomes ``heavier'' (i.e. higher $F_{0.5,1}$ values) at larger redshift for \lom{}. However, since $F_{0.5,1}$ is rather insensitive to the high-mass IMF slope, the correlation remains flat at the Chabrier value for \him{} at all redshifts. 

The result from \citet{Martin-Navarro2015a} is shown in \Fig{F05_vs_z} as grey diamonds, which agrees well with the trend for \lom{} at $z=1$. Indeed, the magnitude of the weak evolution seen in \citet{Martin-Navarro2015a} from $z=0$ to 1 is reproduced well by \lom{}. This good agreement is quite interesting given the fact that \citet{Martin-Navarro2015a} model the increased dwarf fractions by steepening the high-mass slope of the IMF, rather than the low-mass slope as is done in our LoM model. This result highlights the degeneracy between the low- and high-mass slopes in setting the dwarf-to-giant ratio derived in spectroscopic IMF studies of old ETGs.

While of our two variable IMF models, only \lom{} is able to reproduce observational IMF trends based on the dwarf-to-giant ratio, \him{} alone is consistent with those based on the H$\alpha$ flux of local star-forming galaxies (see Paper I). High-redshift observations of star-forming galaxies also find evidence for shallow high-mass IMF slopes in such systems \citep[e.g.][]{Nanayakkara2017, Zhang2018}. In particular, \citet{Nanayakkara2017} find that the enhanced H$\alpha$ equivalent widths (EW) of $z\approx 2$ galaxies could be explained with an IMF slope shallower than (under our IMF definition) $-2.0$. To compare with their data, we compute the ratio of ionizing flux to the SDSS $r$-band flux, $f_{\rm ion}/f_{r}$, for our simulated galaxies, where $f_{\rm ion}$ is the flux of photons with $\lambda < 912 \buildrel _\circ \over {\mathrm{A}}$, and is a proxy for the H$\alpha$ flux \citep{Shivaei2018}. Note that $f_{\rm ion}$ includes light from stars of all ages, while $f_{r}$ is only computed for those with age $> 10$ Myr. The ratio $f_{\rm ion}/f_r$ is then a rough proxy for H$\alpha$ EW, which is the H$\alpha$ flux relative to the continuum. To remove systematics in converting from this quantity to H$\alpha$ EW, we normalize by the value expected given a Salpeter IMF, making it the ``excess'' $f_{\rm ion}/f_{r}$. Since in Paper I we found that the high-mass IMF slope correlates strongly with SFR surface density (a result of the pressure dependence of the IMF), in the upper panel of \Fig{highmassslope_and_HalphaEW_vs_z} we plot the excess $f_{\rm ion}/f_{r}$ as a function of Salpeter-reinterpreted SFR surface density $\Sigma_{\rm SFR,Salp}$, at different redshifts. $\Sigma_{\rm SFR,Salp}$ is computed as in Paper I, where $\Sigma_{\rm SFR,Salp} = {\rm SFR_{Salp}}/(2\pi r_{\rm e,FUV}^2)$, where SFR$_{\rm Salp}$ is the total SFR within a 3D aperture of radius 30 pkpc multiplied by the ratio between the GALEX FUV flux and that expected given a Salpeter IMF, and $r_{\rm e,FUV}$ is the half-light radius in the FUV band. Here we show only star-forming galaxies (intrinsic $u^*-r^*<2$) with $M_r<-19.5$, for consistency with the selection of \citet{Gunawardhana2011}. 

In the upper panel of \Fig{highmassslope_and_HalphaEW_vs_z} we see that the excess $f_{\rm ion}/f_{r}-\Sigma_{\rm SFR,Salp}$ relation is positive for star-forming galaxies in \him{} but flat for those in \lom{}. It also does not evolve with redshift, but is instead extended to larger $\Sigma_{\rm SFR, Salp}$, and thus to larger excess $f_{\rm ion}/f_{\rm r}$. We compare with results from the ZFIRE survey by \citet{Nanayakkara2017}, shown as blue stars. We use their SFRs derived from the H$\alpha$ luminosity, which, due to its sensitivity to SFR on very short time scales, is most consistent with our simulated instantaneous SFRs. For their galaxies we compute $\Sigma_{\rm SFR,Salp}$ as ${\rm SFR}/(2 q \pi r_{\rm e}^2)$, where $r_e$ is obtained by cross-matching the ZFOURGE galaxies with CANDELS and using sizes from \citet{vanderWel2012}, and $q=b/a$ is the minor-to-major axis ratio. While they find a trend of increasing excess H$\alpha$ EW with increasing $\Sigma_{\rm SFR,Salp}$ (in qualitative agreement with \him{} albeit with large scatter), their values tend to be lower than those of \him{}, suggesting that our IMF variations may be too strong in the HiM model. 

It is also interesting that the ionizing excess falls below the Salpeter value for some ZFIRE galaxies with $\Sigma_{\rm SFR} < 1\,\Msun\,{\rm yr}^{-1}\,{\rm kpc}^{-2}$. It is tempting to infer from this result that IMF high-mass slopes steeper than Salpeter are required at lower pressures. However, such low values may also result in the case that these galaxies are post-starburst systems. In this case, the long-lived stars created in the starburst event serve to enhance the continuum more than the H$\alpha$ flux, leading to lower H$\alpha$ EWs than expected for a constant star formation history, even with a fixed IMF (see Fig. 17 of \citealt{Nanayakkara2017}). Thus, stronger constraints on their star formation histories would be required to make inferences on the IMF slope in this regime.

To aid comparison with observations of high-redshift star-forming galaxies, we show in the lower panel of \Fig{highmassslope_and_HalphaEW_vs_z} the high-mass slope of the IMF as a function of $\Sigma_{\rm SFR,Salp}$ at different redshifts. For reference we also show the IMF slope at $z \approx 0.1$ for star-forming galaxies from the GAMA survey inferred by \citet{Gunawardhana2011} assuming either \citet{Calzetti2001}/\citet{Cardelli1989} or \citet{Fischera2005} dust corrections (dust1 and dust2, respectively). As for the ionizing fluxes, this relation does not evolve with redshift for the simulated galaxies, but shifts to higher $\Sigma_{\rm SFR,Salp}$, making the average high-mass slope shallower in high-$z$ star-forming galaxies. These slopes are shallower than those found for the $z\approx2$ observations by \citet{Nanayakkara2017}, for which high-mass IMF slopes of $\approx -2$ may be needed to explain their H$\alpha$ EWs. \citet{Zhang2018} recently concluded that high-mass IMF slopes as shallow as $-2.1$ would be required to explain the ${\rm ^{13}CO/C^{18}O}$ ratios observed in $z\approx 2-3$ submillimetre galaxies, which is also steeper than predicted by \him{}. These findings suggest that the high-mass slope may be too shallow in our \him{} simulation.

\subsection{ Redshift-dependent cosmic properties }
\label{sec:redshift_cosmic}

%Fig 15 - SFR density and Mass density vs redshift
\begin{figure}
  \centering
\includegraphics[width=0.45\textwidth]{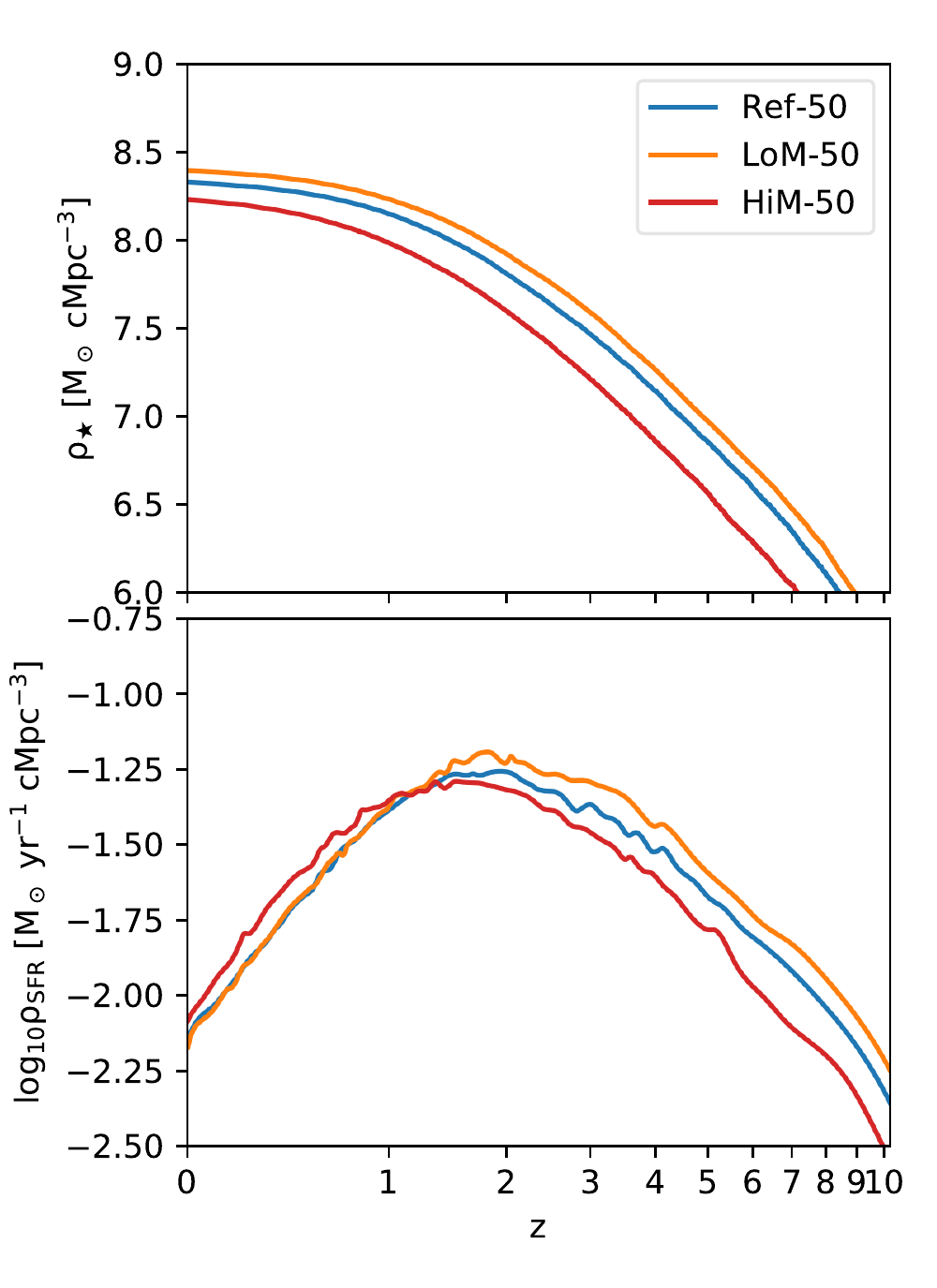}
\caption{Comoving cosmic stellar mass density (top panel) and SFR density (bottom panel) as a function of redshift in the Ref-50 (blue), \lom{} (orange) and \him{} (red) simulations. Star formation is enhanced and suppressed at high $z$ in \lom{} and \him{}, respectively. }
\label{fig:cosmic_evolution}
\end{figure}

We now briefly investigate the redshift evolution of star formation in the simulations. The upper panel of \Fig{cosmic_evolution} shows the evolution of the cosmic comoving stellar mass density, $\rho_\star$, in our variable IMF simulations compared with Ref-50. At all redshifts, $\rho_\star$ lies systematically above and below Ref-50 in \lom{} and \him{}, respectively. The offsets are stronger at high redshift and gradually come into closer agreement with Ref-50 with time, where from $z=5$ to $z=0$ the offset for \lom{} drops from $\approx 0.1$ to 0.07 dex, and that for \him{} drops from $-0.3$ to $-0.1$ dex. The stronger offsets at high redshift are likely due to the fact that the stars that formed at the highest pressures tend to have formed earlier, while those with more Chabrier-like IMFs typically formed later. Note as well that here we plot the stellar masses directly from the EAGLE simulation, with no post-processing with FSPS. Since FSPS assumes higher remnant masses for metal-rich stellar populations than the \citet{Wiersma2009b} models built into EAGLE, $\rho_\star$ shown here for \him{} may be lower than that that would be derived using FSPS. 

These offsets can be explained with the cosmic SFR density, $\rho_{\rm SFR}$, shown in the lower panel of \Fig{cosmic_evolution}. Note that here we show the true values, rather than those that would be inferred assuming a fixed IMF. For \lom{}, $\rho_{\rm SFR}$ is consistent with Ref-50 at low redshift, but at $z>1.5$ it lies systematically $\approx 0.1$ dex above the Ref-50 relation. This offset is likely due to the increased SFR from the star formation law renormalization that is required observationally (see Paper I), since at high redshift, most star formation occurs at high pressure, leading to bottom-heavy IMFs and thus a higher normalization of the star formation law. 

For \him{}, the opposite is true, where at high redshift the SFR is lower than Ref-50 by $\approx 0.1$ dex. This could be due to the renormalization of the star formation law as well (which is decreased at high pressure in \him{}), and/or due to the stronger stellar feedback associated with shallow high-mass IMF slopes. Stronger feedback would be more effective at keeping gas out of galaxies, delaying its reaccretion to later times. This hypothesis is supported by the fact that $\rho_{\rm SFR}$ is larger by $\approx 0.1$ dex at $z<1.5$, and the fact that the cold gas fractions are larger by $\approx 0.5$ dex in high-mass galaxies in \him{} relative to Ref-50 and \him{} (see Fig. 12 of Paper I).

We showed in Paper I that the $K$-band luminosity functions of Ref-50, \lom{}, and \him{} are all consistent with observations at $z \approx 0$. We investigated the evolution of this relation for our simulations, finding no significant difference between them. However, the poor statistics at the bright end of the luminosity functions in our (50 Mpc)$^3$ boxes preclude a detailed comparison, since it is at the bright (high-mass) end where the IMF is expected to have the largest impact. Simulations with larger volumes would thus be required to make a proper comparison.

\section{Discussion}
\label{sec:Discussion}

Throughout this project (Papers I, II, and this work), we have tested a model of galaxy formation and evolution that self-consistently includes local IMF variations that have been calibrated to reproduce the observed MLE$-\sigma_e$ relation of \citet{Cappellari2013a}. In this section we discuss some of the lessons learned and propose some steps for the future.

To make the simulations self-consistent, we made the stellar feedback, star formation law, and the metal yields depend on the IMF. Ensuring that each of these processes was self-consistent was vital in capturing the full effect of the IMF variations. Interestingly, this resulted in galaxy observables such as the luminosity function and BH masses to be mostly unchanged relative to a model with a fixed Chabrier IMF.

As discussed in Papers I and II, this is partially because while the change in IMF modifies the galaxies' masses through its effect on the stellar feedback, for the luminosity function this effect is largely cancelled by the modified $M/L$ ratios resulting from the IMF variations. This story is, however, further complicated by the fact that in the mass regime in which the IMF varies significantly from the Chabrier form ($\Mstar \gtrsim 10^{10.5}\Msun$) both stellar and AGN feedback play important roles in regulating the inflow of gas onto the galaxy. In this regime one might expect BH feedback (and thus BH growth and the $z=0$ BH masses) to adjust to account for the modified stellar feedback resulting from IMF variations. However, we found that our modifications to make the star formation law self-consistent precluded this (likely by adjusting the gas densities near the BHs), leading to similar BH masses as in the reference model. Thus, when performing variable IMF simulations, it is very important to include both self-consistent prescriptions for stellar feedback and the star formation law to ensure that the full effect of the IMF variations is captured and a realistic population of galaxies is produced.

For other galaxy properties, such as the metallicity and alpha enhancement, we have shown that the choice of IMF parametrization is extremely important when implementing IMF variations into galaxy formation models. Indeed, even though the LoM and HiM models produced similar MLE-$\sigma_e$ relations, they resulted in very different metallicities and alpha abundances because the high-mass slope is especially important in setting the feedback strength and metal enrichment of galaxies. We thus encourage observational studies that infer the IMF slope to be careful in choosing which part of the IMF they should vary. For example, \citet{LaBarbera2013} use the effect on the dwarf-to-giant ratio imprinted onto the spectra of old ETGs to constrain the IMF slope at all masses (0.1 to 100 $\Msun$) or only at high masses. However, we have shown that the dwarf-to-giant ratio can be increased simply by varying the IMF only at low masses, leaving any variation of the high-mass IMF unconstrained by spectroscopic measurements of (old) ETGs. Indeed, we argue that varying only the low-mass slope is much preferred over the high-mass slope due to the strong effect the latter has on metal enrichment and stellar feedback in galaxy formation models, making the former the simplest model that explains the observations (see also \citealt{Martin-Navarro2016}).

Despite the success of our variable IMF simulations in terms of reproducing observationally-inferred global and radial trends, we have shown that they are unable to simultaneously reproduce the inferred dwarf-to-giant ratios in ETGs, which prefer LoM, as well as the ionizing properties of late-type galaxies, which prefer HiM. In order to match all of these observations simultaneously, a more complex model of IMF variations, such as a hybrid of the LoM and HiM models, may be required. Indeed, the shallow high-mass slope in the HiM model was required to obtain sufficiently large MLE values from stellar remnants to reproduce the MLE$_r-\sigma_e$ relation of \citet{Cappellari2013b} when the low-mass slope was kept fixed. If the low-mass slope were allowed to vary in tandem with the high-mass slope, such shallow high-mass slopes would no longer be required to obtain large MLE values, as the MLE would then increase due to a combination of excess dwarf stars {\it as well as} excess stellar remnants. Such a model would need to become simultaneously bottom-heavy and top-heavy towards higher-pressure environments, possibly by steepening and levelling the low-mass and high-mass IMF slopes, respectively. This ``butterfly'' IMF model would need to be calibrated to simultaneously reproduce the MLE and dwarf-to-giant ratios of high-mass ETGs, as well as the ionizing fluxes of strongly star-forming galaxies.  A similar model has already been implemented into semi-analytic models of galaxy formation with promising potential to reconcile these observations \citep{Fontanot2018}, and would thus be interesting to implement into our self-consistent hydrodynamical, cosmological simulations.
 
\section{Summary and Conclusions}
\label{sec:conclusions}

In Paper I we presented two cosmological, hydrodynamical simulations based on the EAGLE model that self-consistently vary the IMF to become either more bottom-heavy (\lom{}) or top-heavy (\him{}) locally in high-pressure environments. These IMF prescriptions were calibrated to match the observed relation between the ``galaxy-wide'' excess mass-to-light ratio relative to a fixed IMF (the MLE) as a function of stellar velocity dispersion, by respectively increasing the mass fraction of dwarfs or stellar remnants in high-pressure environments (\Fig{IMF}). In Paper II we showed how the MLE varies globally between galaxies as a function of various galaxy properties, including metal enrichment, age, and BH mass. In this paper, we investigate radial trends within the galaxies in these simulations to compare with the EAGLE simulations with a universal, Chabrier IMF (Ref-50) and with observed radial gradients of IMF diagnostics and abundances in high-mass galaxies, in addition to investigating the redshift dependence of the IMF in our simulations. 

In order to investigate internal trends that could be compared in a meaningful way with observations, we select all galaxies with intrinsic $u^*-r^*>2$ and $\sigma_e > 150\kms$ from each simulation, hereafter referred to as the ``Sigma150'' sample (\Fig{sample}). Our conclusions regarding trends within these galaxies are as follows:
\begin{itemize}
 \item 
 High-$\sigma_e$ galaxies exhibit strong radial IMF gradients, being either bottom- or top-heavy in the centre for \lom{} and \him{} galaxies, respectively, and gradually transitioning to a Chabrier/Kroupa-like IMF beyond the half-light radius (Figs. \ref{fig:alpha_maps}, \ref{fig:IMF_vs_r_sigmabins} and \ref{fig:IMFslope_vs_r}). These trends result from the inside-out growth of galaxies, where the stars in central regions tend to have formed earlier from interstellar gas with higher pressures than the stars in the outer regions (\Fig{age_pressure_vs_r}).
 \item
 For both simulations the IMF gradients result in an increase in the MLE toward the central regions, in qualitative agreement with recently observed MLE gradients in high-mass ETGs. While all high-$\sigma_e$ galaxies in \lom{} display negative MLE gradients, \him{} galaxies exhibit more diverse behaviour, with radial MLE gradients ranging from strongly negative to flat (\Fig{MLE_vs_r}). When measuring the IMF via the fraction of mass or light coming from low-mass dwarf stars, \lom{} is consistent with observations of the centres of high-mass ETGs, while \him{} is in tension with the data (\Fig{fdwarf_vs_r}).
  \item
 The difference in the radial MLE behaviour between \lom{} and \him{} is likely a result of the strong dependence of the MLE on age in \him{}, even at fixed IMF slope, as well as the larger scatter in birth ISM pressure within $r_e$ in \him{} galaxies (\Fig{age_pressure_vs_r}).
 \item
 All of our simulations produce negative stellar metallicity gradients, in agreement with observations. \lom{} and Ref-50 produce weakly positive stellar [Mg/Fe] gradients. However, \him{} has enhanced [Mg/Fe] within $r_e$, causing [Mg/Fe] gradients to be much flatter, which may be more consistent with observed flat [Mg/Fe] profiles in high-mass ETGs \citep[e.g. ][]{Mehlert2003, Parikh2018}. While $M/L$ gradients are flat in Ref-50, \lom{} galaxies have strong negative $M/L$ gradients and \him{} galaxies show a large diversity in $M/L$ gradients, ranging from strongly positive to strongly negative (\Fig{abundances_vs_r}). Such gradients must be taken into account when making inferences on the IMF via dynamical masses.
 \item
 We find strong positive correlations between the local MLE and local stellar metallicity for both \lom{} and \him{} galaxies, qualitatively consistent with observations (\Fig{IMF_vs_abundances}, top row). This is true when considering all galaxies simultaneously or for individual galaxies. Correlations between MLE and local metallicity can thus occur naturally even when the IMF is not governed by metallicity at all.
 \item
 We find strong negative and positive correlations between the local MLE and local stellar [Mg/Fe] for \lom{} and \him{}, respectively, when considering all galaxies simultaneously (\Fig{IMF_vs_abundances}, middle row). This finding may aid in distinguishing between these two IMF variation scenarios with observations.
 \item
 When considering all Sigma150 galaxies simultaneously, the local MLE shows a weak positive correlation with the local stellar age for \lom{} (\Fig{IMF_vs_abundances}, bottom row). However, the local MLE shows no systematic correlation with local age for individual galaxies in either simulation, implying that the overall local MLE$-$age correlations seen for the whole Sigma150 population are a global, rather than a local, property, consistent with the strong positive global MLE$-$age relation reported in Paper II. It is thus important for studies of the spatially-resolved IMF to remove global trends between the MLE and galaxy properties before making inferences on local ones.
  
\end{itemize}

We have investigated the redshift dependence of the IMF and global galaxy properties in our variable IMF simulations. Our results are as follows:
\begin{itemize}
\item
For \lom{}, the MLE$-\sigma_e$ relation has a higher normalization at $z=2$ than at $z=0.1$ due to the typically higher pressures at which stars form at high redshift. In contrast, \him{} produces a much lower normalization at $z=2$ despite the time dependence of stellar birth ISM pressures resulting in more top-heavy IMFs at early times (\Fig{MLE_vs_z}). This lower normalization arises due to the age dependence of the MLE parameter for the HiM prescription (\Fig{MLE_vs_z_particles}).
\item
The dwarf-to-giant fraction increases weakly toward higher redshifts for high-$\sigma_e$ galaxies in \lom{}, consistent with  the observed evolution for quiescent galaxies (\Fig{F05_vs_z}). \him{} is inconsistent with these observations.
\item
The ionizing flux of star-forming galaxies in \him{} increases with star formation rate surface density due to shallower IMF slopes at high pressures, and shows little evolution. This trend is consistent with observed star-forming galaxies, but our IMF slopes may be too shallow relative to the high-redshift observations (\Fig{highmassslope_and_HalphaEW_vs_z}). 
\item
We speculate that a model with a hybrid IMF, which incorporates both a shallow high-mass slope and a steep low-mass slope at high pressures, may be able to simultaneously reproduce the observed ionizing flux in star-forming galaxies as well as the MLE and dwarf-to-giant ratio in ETGs.
\item
At $z>2$, the cosmic SFR density is higher and lower in \lom{} and \him{}, respectively, than in Ref-50. However, below $z \sim 1$, \lom{} matches Ref-50 while \him{} lies above Ref-50. The peak of star formation is shifted toward lower redshift in \him{} (\Fig{cosmic_evolution}). This shift is likely caused by a combination of the decreased star formation law and burstier stellar feedback at high $z$ in \him{}, decreasing the SFR at high $z$ and delaying the infall of cold gas onto galaxies.
\end{itemize}

The findings presented in this paper highlight the importance of being able to spatially resolve the galaxies in which the IMF is constrained. Indeed, radial IMF gradients can have significant implications for dynamical mass measurements used to constrain the IMF \citep[e.g.][]{Bernardi2018b}, as well as when comparing quantities measured within different apertures \citep[see][]{vanDokkum2017}. We have also shown that correlations between the IMF and local galaxy properties can differ both qualitatively and quantitatively from global relations between the same quantities. Thus, studies that do spatially resolve the IMF should take care to avoid confusing global and local IMF scaling relations when combining results for different galaxies.

\section*{Acknowledgements}

We are grateful to the anonymous referee for improving the quality of this work. We thank Madusha Gunawardhana for helpful discussions and Themiya Nanayakkara for providing data from ZFIRE. This work used the DiRAC Data Centric system at Durham University, operated by the Institute for Computational Cosmology on behalf of the STFC DiRAC HPC Facility (www.dirac.ac.uk). This equipment was funded by BIS National E-infrastructure capital grant ST/K00042X/1, STFC capital grants ST/H008519/1 and ST/K00087X/1, STFC DiRAC Operations grant ST/K003267/1 and Durham University. DiRAC is part of the National E-Infrastructure. RAC is a Royal Society University Research Fellows. We also gratefully acknowledge PRACE for awarding us access to the resource Curie based in France at Tr$\grave{\rm e}$s Grand Centre de Calcul. This work was sponsored by the Dutch National Computing Facilities Foundation (NCF) for the use of supercomputer facilities, with financial support from the Netherlands Organization for Scientific Research (NWO). This research made use of {\sc astropy}, a community-developed core {\sc python} package for Astronomy \citep{Astropy2013}. This work has benefited from the use of Py-SPHViewer \citep{Benitez-Llambay2017}.

%%%%%%%%%%%%%%%%%%%%%%%%%%%%%%%%%%%%%%%%%%%%%%%%%%



\bibliographystyle{mnras} % This defines the style of the bibliography
\bibliography{IMF} % This is the file with all the references

%%%%%%%%%%%%%%%%%%%%%%%%%%%%%%%%%%%%%%%%%%%%%%%%%%

% Don't change these lines
\bsp	% typesetting comment
\label{lastpage}
\end{document}